\documentclass[journal,draftclsnofoot,onecolumn]{IEEEtran}
\usepackage{cite}
\usepackage{url}
\usepackage{amsmath,amssymb,amsfonts}
\usepackage{algorithmic}
\usepackage{graphicx}
\usepackage{subfigure}
\usepackage{paralist}
\usepackage{listings}
\usepackage[T1]{fontenc}
\usepackage[utf8]{inputenc} 
\usepackage{makecell}
\usepackage[dvipsnames]{xcolor}
\usepackage{color}
\usepackage{ulem}
\definecolor{maroon}{rgb}{0.5,0,0}
\definecolor{darkgreen}{rgb}{0,0.5,0}
\lstdefinelanguage{XML}
{
  basicstyle=\ttfamily,
  morestring=[b]",
  moredelim=[s][\bfseries\color{maroon}]{<}{\ },
  moredelim=[s][\bfseries\color{maroon}]{</}{>},
  moredelim=[l][\bfseries\color{maroon}]{/>},
  moredelim=[l][\bfseries\color{maroon}]{>},
  morecomment=[s]{<?}{?>},
  morecomment=[s]{<!--}{-->},
  commentstyle=\color{darkgreen},
  stringstyle=\color{blue},
  identifierstyle=\color{red},
  tabsize=1
}
\usepackage{textcomp}
\usepackage{tikz}
\usetikzlibrary{shapes}

\def\BibTeX{{\rm B\kern-.05em{\sc i\kern-.025em b}\kern-.08em
    T\kern-.1667em\lower.7ex\hbox{E}\kern-.125emX}}

\DeclareMathOperator*{\argmin}{arg\,min}

\newcommand{\define}[1]{\textbf{#1}}

\newcommand{\vect}[1]{\ensuremath{\mathbf #1}}

\newcommand{\reals}{{\ensuremath{\mathbb R}}}

\newcommand{\set}[1]{\ensuremath{\left\{ #1 \right\}}}
\newcommand{\setof}[2]{\ensuremath{\left\{ #1 \middle| #2 \right\}}}

\newcommand{\fnct}[1]{\ensuremath{#1(\cdot)}}

\newcommand{\inner}[2]{\ensuremath{\left\langle #1, #2 \right\rangle}}
\newcommand{\norm}[1]{\ensuremath{\left\|#1\right\|}}

\newcommand{\parconv}[3]{{\ensuremath{\sigma_{{#1}|{#2}}({#3})}}}

\usepackage{tikz}
\usetikzlibrary{automata, positioning, arrows}
\tikzset{
  ->, 
  >=stealth', 
  node distance=3cm, 
  every state/.style={thick, fill=gray!10}, 
  initial text=$ $, 
}

\makeatletter
\def\Expect{\mathop{\operator@font E}\nolimits}
\makeatother

\newcommand{\magnitude}[1]{\ensuremath{\left\lvert #1 \right\rvert}}

\newcommand{\vpref}{{\ensuremath{v_{\mbox{pref}}}}}
\newcommand{\vmax}{{\ensuremath{v_{\mbox{max}}}}}
\newcommand{\vmin}{{\ensuremath{v_{\mbox{min}}}}}
\newcommand{\candlanes}[2]{{\ensuremath{\mathcal{C}(#1,#2)}}}
\newcommand{\neighlanes}[2]{{\ensuremath{\mathcal{N}(#1,#2)}}}
\newcommand{\blockinglanes}[2]{{\ensuremath{\mathcal{B}(#1,#2)}}}
\newcommand{\dlane}[1]{{\ensuremath{\mbox{\textsc{lane}}(#1)}}}
\newcommand{\dpos}[1]{{\ensuremath{\mbox{\textsc{pos}}(#1)}}}
\newcommand{\dspeed}[1]{{\ensuremath{\mbox{\textsc{speed}}(#1)}}}
\newcommand{\drones}[2]{{\ensuremath{\Delta(#1,#2)}}}
\newcommand{\dronesunion}[1]{{\ensuremath{\Delta(#1)}}}

\newcommand{\distahead}[2]{{\ensuremath{d^{\mbox{ahead}}_{#1,#2}}}}

\newcommand{\numnearbysame}[2]{{\ensuremath{n^{\mbox{same}}_{#1,#2}}}}
\newcommand{\numnearbyneigh}[2]{{\ensuremath{n^{\mbox{neigh}}_{#1,#2}}}}
\newcommand{\Dronesnearbysame}[2]{{\ensuremath{\Delta^{\mbox{same}}_{#1,#2}}}}
\newcommand{\Dronesnearbyneigh}[2]{{\ensuremath{\Delta^{\mbox{neigh}}_{#1,#2}}}}
\newcommand{\hopdistance}[2]{{\ensuremath{\left[#1,#2\right]_1}}}
\newcommand{\closingpoints}[2]
{{\ensuremath{\mbox{\textsc{closingpoints}}(#1,#2)}}}
\newcommand{\ramppoints}[2]{{\ensuremath{\mbox{\textsc{ramppoints}}(#1,#2)}}}
\newcommand{\dronepoints}[2]{{\ensuremath{\mbox{\textsc{dronepoints}}(#1,#2)}}}
\newcommand{\aheadpoints}[2]{{\ensuremath{\mbox{\textsc{aheadpoints}}(#1,#2)}}}
\newcommand{\closestaheaddist}[2]{{\ensuremath{\mbox{\textsc{minaheaddist}}(#1,#2)}}}
\newcommand{\behindpoints}[2]{{\ensuremath{\mbox{\textsc{behindpoints}}(#1,#2)}}}

\newcommand{\closestbehinddist}[2]{{\ensuremath{\mbox{\textsc{minbehinddist}}(#1,#2)}}}

\begin{document}

\title{Short-Term Guidance Algorithm on a Drone Road System}


\author{\IEEEauthorblockN{ Zhouyu Qu, Andreas Willig, Xiaobing Wu\textsuperscript{\textdagger}}\\
\IEEEauthorblockA{\textit{Department of Computer Science and Software Engineering}\\ \textsuperscript{\textdagger}\textit{Wireless Research Centre} \\
\textit{University of Canterbury}\\
Christchurch, New Zealand \\
zhouyu.qu@pg.canterbury.ac.nz, \{andreas.willig, barry.wu\}@canterbury.ac.nz}
}

\maketitle



\begin{abstract}
  Unmanned Aerial Vehicles (UAVs), commonly known as drones, have
  experienced expanding use in urban environments in recent
  years. However, the growing density of drones raises significant
  challenges, such as avoiding collisions and managing air traffic
  efficiently, especially in congested areas. To address these issues,
  a structured road system and an effective guidance algorithm are
  essential. In this paper, we introduce a markup language allowing to
  describe drone road systems (DRS), in which a road system is given
  by a set of individual roads, each of which can have a varying
  number of lanes. Roads can be linked through connecting
  lanes. Furthermore, we propose a novel short-term decentralized
  greedy (STDG) guidance algorithm that uses only the position and
  speed information of nearby drones — communicated via periodically
  transmitted beacons — to make real-time decisions such as stopping,
  changing lanes, or adjusting speed for the next few seconds. Unlike
  existing methods that rely on centralized coordination, our
  algorithm enables drones to operate independently while ensuring
  safety and efficiency. We present simulation results showing the
  impact of key wireless and algorithm parameters on performance
  metrics like the drone collision rate, average speed and throughput
  of the drone road system.
\end{abstract}


\begin{IEEEkeywords}
  Drone, Congestion Control, Path-Finding, Road System, Wireless Communication
\end{IEEEkeywords}

\section{Introduction}
\label{sec:intro}

With economic development, ground vehicles have become an affordable
and popular tool for improving people's quality of life. As of 2017,
over ninety percent of adults in the U.S. have used cars to commute to
work\cite{moody2021value}. However, the increasing number of vehicles
presents several critical problems, including environmental problems
\cite{johnson2017cars} and the exhaustion of natural resources
\cite{piotrowska2019assessment}. Moreover, traffic congestion has
become a significant challenge that affects various sectors, including
healthcare and delivery of packets and other goods, where timely
transportation is crucial.

Drones provide a compelling alternative to ground vehicles for
addressing some of these transportation issues, in particular the
delivery of small items such as medicine or
pizza\cite{dukkanci2024facility}.  Unlike traditional vehicles, drones
can navigate in three dimensions, enabling them to select more
efficient routes and avoid traffic congestion. Their ability to fly in
the air grants them greater freedom to bypass obstacles on the
ground. Moreover, drone-based delivery is potentially more
environmentally sustainable\cite{li2023drone}, as the mass of a drone
is significantly lower than that of a conventional vehicle. This makes
drones an interesting option for transporting small goods, especially
in densely populated urban areas. Recently, interest in using drones
for deliveries has
surged\cite{eskandaripour2023last,rajabi2022drone,qu2022sensorless},
and it is anticipated that they will become a common trend for parcel
delivery in the near future. However, drones have limited carrying
capacities, typically ranging from 0.3 to 20
kg\footnote{\url{https://www.droneblog.com/drone-payload/} (accessed
  on 18 March 2025).}. To meet the demand for current delivery volumes
-- such as approximately 2.3 million
packages\footnote{\url{https://www.crainsnewyork.com/technology/inside-new-york-citys-wild-west-package-delivery}
  (accessed on 18 March 2025).} delivered daily in a city like New
York -- a large fleet of drones will be necessary. This introduces a
significant challenge: the risk of crashes. With hundreds or even
thousands of drones operating in the same airspace at the same time,
the likelihood of collisions dramatically increases if no coordination
is introduced. Without clearly defined airspace restrictions, the
environment can become chaotic, with each drone acting as an obstacle
and a potential hazard to others. As each drone travels at different
speeds and in various directions, navigation becomes increasingly
unpredictable and complex. Consequently, drones may struggle to avoid
collisions and have to fly at lower speeeds or require elaborate
maneuvers to avoid other flying objects.

One approach to manage the large population of drones is to limit
their movements to specific areas which collectively are called a
\define{drone road system} (DRS). In our conception, such a system
should mimic essential ground transport features such as lanes, roads,
and ramps. Drone cruising within the system should comply with some
rules to restrict their movements. Unlike the ground road system for
cars, a DRS is pre-planned in 3-D airspace, it does not require major
construction activities and is therefore more easily deployable than
their ground-based counterparts.

This paper makes two main contributions. The first is a design for a
machine-readable way to describe a drone road system. In particular,
we have developed an XML-based representation which will later on
allow for the design of electronic support tools such as editors,
automatic validity checks, and so on. Our XML representation involves
the following key elements:
\begin{inparaenum}[(i)]
\item A \define{drone road} is described by an oriented smooth spatial
  curve and all drones on this road move in the same direction
  following its orientation;
\item A drone road is sub-divided into a number of parallel
  non-overlapping \define{lanes} such that drones normally move within
  one lane most of the time (except when switching lanes);
\item A \define{ramp} is used either for entering or leaving a drone
  road (more precisely: one of its lanes), in which case they are
  called \define{on-ramp} or \define{off-ramp}, or a ramp
  interconnects two (lanes of) two different roads, in which case the
  ramp is called a \define{connecting ramp}. Drones enter and leave
  the drone road system exclusively through on- or offramps, and they
  switch roads through connecting ramps.
\end{inparaenum}
All these elements are described in three dimensions, using a
pre-arranged coordinate system, which for the purposes of this paper
we assume to be a standard Cartesian $x/y/z$ coordinate system with a
known reference point and orientation with respect to the earth
surface. In this paper we assume that the XML representation of a
drone road system is available to and consistent across all involved
drones and actually represents a feasible road system design. Note
that details about how the DRS description is conveyed to the drones,
or how the feasibility of a road system is checked (e.g.\ ensuring
that roads and lanes do not intersect or have sufficiently small
curvature to allow drones to move at high speeds) are outside the
scope of this paper.

The second contribution is a guidance algorithm for individual
drones. We assume that the end-to-end path of a drone that leads it
from its starting point to its destination is pre-calculated, for
example from executing a shortest-path-type algorithm outputting the
sequence of on-ramp, drone roads and connecting ramps, and off-ramp
the drone will have to take. Once the drone has entered the DRS, it
will need to continuously make decisions about its speed and the lane
to travel on, ideally in such a way that the average speed of the
drone is as close to its preferred speed as possible while collisions
with other drones moving along the same road or lane are avoided. We
refer to the decision-making entity for these shorter-term decisions
as the drone's \define{guidance system} or \define{guidance
  algorithm}. The design of the guidance algorithm will have
significant impact on key performance parameters of a drone road
system: the (average) number of drone collisions, the average speed
that drones can achieve, and the throughput of the system (related to
the number of drones reaching their final destination per unit
time). These performance parameters have complex interactions and
tradeoffs, which we begin to explore in this paper through a
simulation study.

We hope that once the concept of a drone road system has gained some
traction, there will be substantial research in guidance
algorithms. In this paper we propose a guidance algorithm which we
intend to be a reference algorithm for further designs. One key reason
for us regarding our algorithm as more of a baseline or reference
algorithm rather than an optimal solution with (close-to) optimal
performance is that we explicitly and deliberately make only minimal
assumptions about the data available to the guidance algorithm. In
particular, accepting that in a general DRS the drones will be of
quite variable capability, ownership, build and maintenance level, we
do not make assumptions about universally available on-board sensors
(like for example a LIDAR), and we only assume that all drones have a
GPS receiver of sufficient quality and a wireless interface (like WiFi
or C-V2X) available. Through this wireless interface drones exchange
safety data with neighboured drones, e.g.\ each drone will frequently
broadcast its own position, speed and heading, and they will adjust
their own speed and lane based on safety data overheard from
neighboured drones and their given end-to-end path. These safety data
items are transmitted periodically, we refer to the data packets
containing the safety data as \define{beacons}.\footnote{Note that
  this is very similar to the approach taken in vehicular networks,
  which also critically relies on periodically transmitted beacons
  with safety data to estimate the trajectory of other vehicles and
  avoid collisions.}  Furthermore, the reference algorithm proposed in
this paper is \define{coordination-free}, by which we mean that drones
make their guidance decision \emph{only} based on the received safety
data and do not attempt to enter active negotiations with other drones
to coordinate their actions. We note that the absence of active
negotiation reduces system complexity and also likely reduces complex
issues around the trust placed in foreign drones entering such
negotiations in good faith -- we assume that it is in the best
interest of every drone operator to publish accurate safety data and
reduce the risk of collisions of their drones. Clearly, our assumption
of only GPS and WiFi being available is a worst-case assumption and
drones having additional sensors will likely be able to make better
guidance decisions. We also made the -- again deliberate -- assumption
that it is indeed drones making guidance decisions based on data
received from immediate neighbours, instead of farming out decisions
onto some sort of infrastructure. While we do regard some level of
infrastructure as unavoidable (e.g.\ to disseminate updates to the
drone road system to drones), we want to retain the flexibility of
drones being able to fly outside the reach of infrastructure nodes.

A key risk of our approach is that the beacon transmissions of
neighboured nodes collide on the wireless channel, leading to packet
losses. The likelihood of this happening depends on a number of
different parameters, including transmit power, beacon rate, drone
density and others. These beacon packet collisions lead to losses of
safety data records and increase the uncertainty about the positions
of other drones.

We analyse the performance of our algorithm based on a series of
simulations using the OMNeT++ discrete-event simulation
framework\footnote{\url{https://omnetpp.org/}} together with the
associated INET simulation
library,\footnote{\url{https://inet.omnetpp.org/}} which offers a
range of pre-defined simulation modules for standard protocols,
including the IEEE 802.11 / WiFi standard. The main performance
metrics we consider are the collision rate, the average speed, and the
throughput, which we define to be the number of drones leaving the
system per unit time. In our experiments, we mainly vary the drone
density, beaconing rate and transmit power, and focus on exploring the
effect of these parameters on our chosen performance metrics.

Our simulation results show that the guidance algorithm can keep
collision rates close to zero when properly configured, while on
average keeping the speed of drones within 5\%–25\% of their preferred
speed. Furthermore, our results indicate that there are tradeoffs
between drone collision rates and achievable (average) speeds.

The remaining paper is structured as follows. In
Section~\ref{sec:related} we introduce related work in the fields of
drone road systems, congestion control for vehicular wireless
networks, and drone guidance algorithms. Then, in
Section~\ref{sec:drs-revised}, we describe our notation for describing
drone road systems. Following this, in Section~\ref{sec:system model}
we clarify the research scope and key assumptions of this paper. In
Section~\ref{sec:algorithm}, we introduce the short-term decentralized
greedy (STDG) guidance algorithm, by which drones periodically make
short-term decisions to ensure safety and efficiency using only
beacons from other drones. Finally, in Section~\ref{sec:results}, we
present and discuss several simulation results for different wireless
communication and algorithm parameters, and in Section
\ref{sec:conclusion}, we conclude the paper and discuss potential
future work.


\section{Related Works}
\label{sec:related}

In this section, we present related studies about drone applications
in city, DRS concept, congestion control in vehicular network, and
related drone short-term path-finding/guidance algorithm.  We will
also briefly introduce our previous study and explain the differences
to this paper.

\subsection{Drone applications in city}
\label{sec:related:droneInCity}

Drones are starting to play an essential role in cities. They
can help in various fields, such as photography \cite{cheng2015aerial,
puttock2015aerial}, infrastructure monitoring \cite{varghese2017power,
flammini2016railway}, medical \cite{rosser2018surgical,
claesson2017time}, or road traffic monitoring \cite{kumar2021novel,
outay2020applications}. However, authorities should also limit their
operations to ensure the safety, privacy, and security of individuals.

In recent years, drones have impacted the lives of ordinary people
through a number of commercial applications. One example is medicine
delivery, for instance vaccines for
COVID-19\footnote{\url{https://www.nature.com/articles/d41591-022-00053-9},
  and
  \url{https://www.weforum.org/impact/drones-delivering-vaccines/}}. Drones
can deliver small and fragile goods to vulnerable populations,
reducing the time required for the handover process and ensuring
timely delivery of goods. Another example is food
delivery\footnote{\url{https://www.technologyreview.com/2023/05/23/1073500/drone-food-delivery-shenzhen-meituan/}}. In
2023, a Chinese city, Shenzhen, began deploying food delivery
drones. People could simply use a mobile app to place an order, and
then pick up food from the designated landing area. Similarly, Amazon
also provides drone delivery services, allowing customers to receive
light packages within one
hour\footnote{\url{https://www.amazon.com/gp/help/customer/display.html?nodeId=T3jxhuvPfQ629BOIL4}}.

Currently, the use of drones in cities is in the preliminary testing
phase. One of the key challenges is how drones can be safely
integrated into the cities, avoiding collisions with buildings,
people, and other drones. The current approaches generally only
consider path-planing and collision avoidance for a small fleet of
drones or drones which under same ownership, in a specific area. There
is a research gap in addressing the safety issues for drones operated
by different entities across an entire city area.  Our research
focuses on providing a general solution to this issue by managing
drones through a standardized road system that can manage drones of
any fleet size and operate in urban areas of any scale, and guiding
them to their destinations safely by a combined decentralized guidance
algorithm.

\subsection{DRS concept}
\label{sec:related:drs}

An important literature about drone roads is a NASA project on drone
traffic management \cite{nasa2015air} which serves as a fundamental
regulation of airspace for drones. It is requested that drones should
not fly higher than 400 ft to minimize the impact of drone operations
on other classes of airspace users
\cite{abeyratne2014convention}. Based on the NASA's restrictions, we
explored five different airspace concepts for drones.

In a NASA concept \cite{jang2017concepts} the airspace is divided into
several layers by altitude. Layers at high altitudes accommodate the
fixed-wing, high speed, cruise flight. Oppositely rotary-wing, low
speed, limited flight should fly in low altitude layers. In each
layer, an airspace structure is situated above the street and between
the tall buildings.

\begin{figure}[htbp]
   \centering
   \includegraphics[scale=0.3]{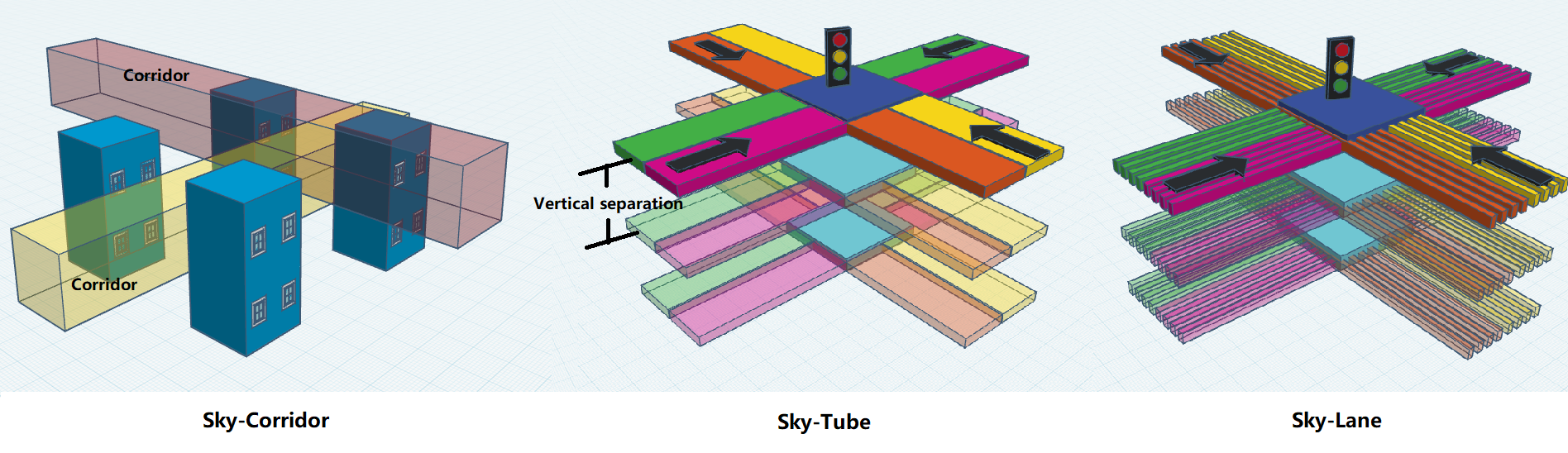}
   \caption{An example of sky-corridors, sky-tube, and sky-lane, from
     \cite{jang2017concepts}}
   \label{fig:lane_tube}
\end{figure}

The concept introduces three airspace structures: sky-tube, sky-lane,
and sky-corridors. An example is illustrated in Figure
\ref{fig:lane_tube}. Sky-corridors are the simplest design, they are
bulk spaces between tall urban buildings. Within the corridor, drones
are allowed to move and change direction freely. Unlike corridors,
drones in the sky-tube can only fly in specific directions and are
unable to fly parallel in a single sky lane. In the design of sky-lane
and sky-tube, traffic lights and time window control systems are
included to help prevent crashes at intersections. Some basic
simulation results using a separation control and cooperative
lane-switching algorithm (introduced in the same paper) are provided
in this research. This shows the shorter period for a traffic light
system can result in a higher drone density, reduced average speed,
and smaller system-wide throughput. However, it does not implement any
simulations to compare the performance of those three airspace
structures.
    
To learn current progress in structured drone traffic management, we briefly review four representative designs, referred to as Concepts I–IV. In Concept I \cite{ijgi10050338}, the authors provide a more specific design of certain parts of the 'drone road', such as a spiral pipe for increasing altitude and a connecting lane for linking two lanes with different direction. These sets of designs are more realistic because they take into account the actual movement trajectory of the drones.
    
In Concept II \cite{aerospace8020038}, M. Doole et al. introduce a
similar drone road concept with NASA's sky-lane. The airspace is also
separated into several layers. They are classified into the through
layers and turn layers. Typically, drones operate in through layers
for cruising, but shift to turn layers when preparing to make a left
or right turn.
    
Concept III is designed by A. Svaigen et
al. \cite{svaigen2022biomixd}. There are only two layers: one is close
to the street and another is a little higher than the urban
building. An intersection is defined as the point when two airway
edges cross with each other, thus in one crossing area, there should
be four intersection points. Drones in this scenario are assumed to
know their own position and velocity information and can send them to
an infrastructure. Then the infrastructure will run algorithms to
calculate the best moving strategy and send it back to the drones.

The last, Concept IV is introduced in \cite{9256627}. It is called
AirMatrix, which is an improved version of the previous concept. The
airspace is divided into many air blocks with different
resolutions. The resolution of the blocks in one area is decided by
the altitude and the type (such as the park, city, or farm) of the
area. In higher resolution areas, the drone flexibility and system
complexity become both high. Note this concept is also designed for
infrastructure involved in drone systems, and communication between
drones is not mentioned in the paper. An adaptive AirMatrix system is
implemented in paper \cite{9945667}. The block size can be adaptively
changed by calculating the GPS signal power and the wind field.

Overall, those concepts can be divided into two classes, layer-based
(includes NASA ones, Concept I, and Concept II) and grid-based
(includes Concepts III and IV). The grid-based designs are high-level
concepts since they just give a grid-like road network, but do not
provide more detail within the road (Do they have lanes? or what does
the intersection look like?). The research gap is, a large fleet of
drones need a fully detailed road design notation for further
algorithm designing. The layer-based designs do not have this
limitation. However, designs except for the concept I just stack some
forms of ground road design, they do not use the superiority of the
3-D airspace (i.e. the lanes in those designs are side by side, and
there is only one route for switching lanes from leftmost to the
rightmost). The concept I gives many details about road parts, but
still lacks the low-level lane design.

\subsection{Congestion Control and Packet Loss in Vehicular Network}
\label{sec:related:congestion}

Vehicular Ad Hoc Networks (VANETs) are the wireless network for the
fast-moving vehicles, mainly addressing the issues of collision
avoidance, traffic flow optimization, and real-time navigation in the
situation which the network topological graph changed very fast, and
low-latency\&high-reliability of packets are highly demanded
\cite{hartenstein2009vanet}.

This research enables the similar network configuration with the
VANETs, which are both based on IEEE 802.11p wireless technology and
periodically transmitted safety beacons by broadcasting to all nearby
nodes. During this process, a key challange is raised, which is called
congestion control \cite{wischhof2005congestion}. When the safety
packets are transmitting, they unavoidable suffer direct or
hidden-terminal collisions with others \cite{fallah2010analysis},
leading to a potential uncertainty for the curial position and speed
information used for node physical collision avoidance. To optimize
the congestion control problem, packet size, transmit power, and
transmission rate are the key factors during the wireless
communication in VANETs \cite{zhang2008congestion}. Related experiment
results \cite{wei2019identifying,1311801,chaabouni2013collision}
indicate that the shorter transmission duration (which equals to the
small size of packets), the lower transmit power and the lower
transmission frequency could reduce the collision rate of
packets. However, to ensure the timeliness and effectiveness of safety
message, these parameter cannot be simply set to a better-side optimal
values. Currently, research on how these parameters affect packet
collisions in the UAV field is limited, and this represents a major
direction for this study.

Packet loss is the key factor that leading the uncertainty of the
safety information, it mainly caused by the packet collision (both
direct and hidden terminal), and the long transmission distance
\cite{gupta2015survey,hota2022performance}. Several literature
contributes on the impact of packet loss on VANETs or drone
networks. In paper \cite{9256561}, several experiments are designed to
explore the how different level of packet loss affect the performance
of three drone path-planning algorithms. Results indicate that with
the increase of the packet loss rate, drone formation suffer more
separation violations, and thus need more buffer area to maintain the
safety needs. Paper \cite{9624939} focuses on exploring the trade off
between packet loss and the performance of UAV swarm localization ,
results show the packet loss, especially due to the long-distance, can
lead to a very promising challenge for localization.

\subsection{Drone Guidance Algorithm}
\label{sec:related:previous}

The guidance/navigation algorithms are commonly used for autonomous
vehicles to continuously update optimal paths, ensuring collision
avoidance with obstacles, buildings, and other road participants while
optimizing energy efficiency and mission success.

Many recent studies have discussed various methods for ground vehicle
to switch lane and avoid collisions. Aksjonov et
al\cite{aksjonov2021rule} introduce a rule-based method to perform
safety turning in a intersection scenario. It defines a general policy
to ensure all vehicles nearby are not potentially risk to ego vehicle,
and set a sub-policy for each neighbor vehicle to evaluate the risk
level. This method is effective in ensuring safety within its specific
scenario, but due to the complexity of real-world traffic situations,
it cannot be directly applied. Hubmann et al\cite{7995949} propose a
method based on POMDP(partially observable Markov decision
process). This method considers multiple possible future movements of
surrounding vehicles and the potential interactions taken from the
current vehicle to others to plan a safe and socially compliant
trajectory. Result shows it can provide a good robustness in an
uncertain environment or some complex scenario, and can also prevent
most incoming collisions. Bea et al \cite{9147837} develop a
cooperation-aware lane change maneuver based on the SGAN(Social
Generative Adversarial Network) for intention prediction and heuristic
algorithm for decision making. This algorithm is mainly designed for
the heavy traffic scenario where lane-switching is much dangerous
without cooperation with other vehicles, and it can output a safe path
based on the predicted movement trajectory of others. Li et al
\cite{9726894} utilize the DRL(Deep Reinforcement Learning) approach,
the trained deep neural network can identify a target lane and decide
whether change to this lane or keep the current lane. This approach is
computationally efficient, making it suitable for the real-time
path-planning.

These guidance algorithms for autonomous ground vehicles are
well-developed to handle continuous path updates, and collision
avoidance during lane switching. However, directly applying these
methods for drone guidance problem is challenging due to several key
differences. The most obvious difference is that drones operate in a
three-dimensional space. In particular, within our proposed road
system structure, drones have more options(lanes) to overtake slower
ones ahead, and they also encounter a greater number of neighboring
drones. This significantly increases the complexity of path planning
and collision avoidance by
\begin{inparaenum}[(i)]
\item requiring much more rules to restrict drone to perform a lane
  switching action in some rule-based algorithm
  \cite{aksjonov2021rule,wang2019lane, xu2022integrated};
\item introducing more uncertainties for the behavior prediction-based
  algorithm \cite{7995949};
\item increasing the complexity of building a learning model for the
  learning-based algorithm \cite{9147837, 9726894}.
\end{inparaenum}
Furthermore, ground vehicles have been widely used for many years,
providing sufficient training data for machine learning and behavior
prediction. In contrast, drones are still at a relatively early stage
of deployment, and then only limited real-world data available for
training and learning. Further differences between ground vehicle and drone guidance also exist. Drones only need to be aware of other flying vehicles, without considering pedestrians, and the drone road system can be easily modified without physical reconstruction, which makes the logic of guidance design fundamentally different. In addition, drones are more sensitive to energy consumption compared to ground vehicles, which suggests that computationally inefficient approaches (e.g.\ Model Predictive Control(MPC) based algorithm \cite{suh2018stochastic}) may not be suitable for drone operations.

Although few studies focus on drone guidance algorithms in scenarios
like ours, some related guidance or collision avoidance algorithms
have been developed for specific situations:
\begin{inparaenum}[(i)]
\item the modified A* algorithm \cite{farid2022modified} or Dijkstra
  \cite{wang2022trajectory} algorithm that treat other drones as
  obstacles and then plan a shortest and safest short-term path for
  drone to avoid collision;
\item the Particle Swarm Optimization (PSO) approach
  \cite{abhishek2020hybrid} that aims to find a global optimal
  solution for collision avoidance, dynamic path planning, and swarm
  behavior management in the drone swarm;
\item Deep Reinforcement Learning (DRL) approach \cite{8820322,
    azar2021drone, hodge2021deep, 9767604} that takes RGB image data
  from camera, and the depth data (i.e.,direction and magnitude of
  obstacle nearby) from few sensors to generate a safe path for their
  specific scenarios.
\end{inparaenum}
 
However, these approaches are not well-suited for environments
characterized by high drone density, limited inter-drone coordination,
and structured aerial traffic systems. In such situations:
\begin{inparaenum}[(1)]
\item drones operate in a crowded airspace;
\item they belong to different owners, making cooperation impractical; and
\item a predefined road structure confines drone movement, meaning
  that all potential collisions originate from other drones.
\end{inparaenum}
As a result, each drone’s trajectory must be computed with extreme
caution to avoid interference with others. Otherwise, that may trigger
a chain reaction and result in a heavy congestion or collisions. these
drones are typically equipped with various sensors, so the algorithm
should run with minimal input data, meaning that the content of input
should be as simple as possible. Based on that, one key contribution
in this study is to propose a coordination-free and decentralized
guidance algorithm which could achieve collision avoidance and
throughput optimization .

In our previous paper\cite{qu2022sensorless}, we proposed a collision
avoidance and lane switching algorithm for a large fleet of
drones. The results indicate that this algorithm could effectively
reduce the collision rate, and both algorithm parameters such as
safety distance and wireless communication parameters can affect the
final result. However, that algorithm only considers a piece of long
straight tube, not a full drone road system which should include
multiple straight and curve roads and intersection.

Building on this foundation, the current work extends the algorithm to
enable its application within a comprehensive drone road system. The
main improvements are:
\begin{inparaenum}[(i)]
\item The shape of the road may include curves, requiring the
  algorithm to determine the positional relationship between drones,
  particularly in high-curvature sections, and to organize suitable
  paths for them to switch lanes.
\item The start and the destination of drones may not be located in
  the same road, requiring the algorithm to deal with the
  road-switching situation.
\item In a high-density drone environment, drones may be unable to
  switch to their target road without waiting. This requires the
  algorithm to organize drones to come to a complete stop behind the
  switch point and facilitate orderly diversion.
\end{inparaenum}
In addition, the scope of the simulation has been extended. The number
of entry points has been increased from a single point to multiple
points, allowing drones to be generated at any of these entries. As a
result, the number of drones in the simulation is expected to increase
significantly, from hundreds to thousands. That provides additional
complexity of designing the algorithm for the congestion control.

Due to the complete stop action mentioned above, throughput becomes an
important metric for the simulation. We will focus on exploring the
effect of the wireless communication and the algorithm parameters on
that metric, and find the optimal values for reducing the complete
stop period to increase the average speed of drones.


\newcommand{\drselem}[1]{\textcolor{blue}{\texttt{#1}}}
\newcommand{\drsattr}[1]{\textcolor{green}{\textit{#1}}}

\section{Drone Road System}
\label{sec:drs-revised}

In this section we propose a XML-based description language for drone
road systems (DRS). We restrict the presentation to key elements and
attributes. Further attributes and elements may exist (for example
things like version numbers, checksums or hashes to check integrity
etc.), but are not relevant for this paper.

\subsection{Overview}

At the lowest level of our DRS formalism we have \drselem{Curve}
elements, which represent sufficiently smooth (at least $C^2$)
mathematical curves $\gamma:[a,b]\mapsto\reals^3$ from a parameter
interval $[a,b]$ into the three-dimensional real space. In a
\drselem{ChainedCurve} element a number of \drselem{Curve} elements
are joined together (or chained together) in a smooth fashion, so that
longer road segments or lanes can be composed out of simpler and
tailored individual curves. A \drselem{Lane} element contains a
\drselem{ChainedCurve} and additionally can assign certain attributes
to each of the \drselem{Curve} elements making up the
\drselem{ChainedCurve}. One example of such an attribute is whether a
curve is open (i.e.\ the \drselem{Curve} can be used by drones) or
closed (drones are not allowed to use the \drselem{Curve}). Drones
travel along lanes, and all drones using the same lane travel in the
same direction.  A \drselem{Road} element contains at least one
\drselem{Lane}, which we also refer to as its \define{center lane},
and it can contain an arbitrary number of further \drselem{Lane}
elements which run in parallel to the center lane, we refer to these
as \define{parallel lanes}. All lanes of a road run in the same
direction, so that in fact all drones on the entire road travel in
that same direction, and roads as a whole are therefore
uni-directional. At the end points of a road drones can enter or leave
the drone road system. To allow a drone to transition from one road to
another or to enter or leave a road at other points than the end-point
of a road, we introduce the \drselem{Ramp} element, which can be
thought of as single-lane directed roads that can either connect from
one particular lane of one road to another particular lane on another
road in a manner that respects directions (such a ramp used to connect
roads is referred to as a \define{connecting ramp}), or which are
attached to one particular lane of a road and then lead to a point
outside the drone road system. These points can be used by drones to
enter or leave a road and correspondingly we call these ramps
\define{onramps} or \define{offramps}. Finally, a
\drselem{DroneRoadSystem} element contains one or more \drselem{Road}
elements and an arbitrary number of \drselem{Ramp} elements.

In all of the following we assume that we are operating in a
pre-established three-dimensional Cartesian coordinate system ($x/y/z$
coordinates) with a known origin and known orientation of its
orthogonal axis vectors. We assume that the (orthogonal) $x$- and
$y$-coordinates lie in a plane tangent to the earth surface at the
origin, and the $z$-axis is orthogonal to these.

\subsection{Curves, Parallel Curves and Chained Curves}

\subsubsection{Curves}

A key feature of our drone road system description method is that it
allows curved lanes and ramps which in turn might be composed or
chained together from multiple curves, making the shapes more flexible
and adaptable, thereby providing much freedom for the road system
designer.

A curve in this paper represents a three-dimensional smooth (at least
$C^2$) mapping from a closed interval $[a,b]\subset\reals$ (with
$a < b$) into $\reals^3$, i.e.\ $\gamma:[a,b] \mapsto
\reals^3$. Hence, a curve has finite length, which is given
by \cite{do2016differential}
\begin{displaymath}
  L(\gamma) = \int_a^b \norm{\gamma'(t))} \,dt
\end{displaymath}
where $\norm{\cdot}$ is the Euclidean norm. A curve $\fnct{\gamma}$
can be understood as modeling the movement of a particle along the
curve as ``time'' passes. We assume that all our curves
$\fnct{\gamma}$ are given in the so-called arc-length representation,
which models the movement of a particle along the curve at unit-speed,
it therefore reflects actual physical travel distances along the
curve. We also assume that the curve is regular, i.e.\ its derivative
$\gamma'(s)$ is non-zero everywhere.

\subsubsection{Parallel Curves}
\label{subsubsec:drs:parallel-curves}

When designing a drone road, we specify a central lane and a number of
lanes parallel to it. To achieve this, we need the ability to
construct curves running in parallel to a given curve
\fnct{\gamma}. The construction of a parallel curve is enabled by the
construction of a moving frame along \fnct{\gamma}, i.e.\ to each
point of \fnct{\gamma} we assign a two-dimensional coordinate system
in the normal plane in this point with certain orthogonal unit vectors
\vect{u_1} and \vect{u_2}, and these unit vectors vary smoothly along
the curve. This construction is known as a moving frame. The parallel
curve is then obtained by picking the same coordinates in the normal
planes for each point \vect{p} of the 'source' curve. A number of
methods are available for constructing a moving frame, and many of
these have restrictions -- for example, the well-known Frenet frame
\cite{bishop1975there} construction will not work when the curve is a
straight line. In Appendix~\ref{sec:app:constructing-parallel-curve}
we describe a way to construct a parallel curve, but we note that our
work is not tied to this and other methods for assigning parallel
curves (such as Euler–Rodrigues Frame \cite{choi2002euler}, and
Rotation Minimizing Frame \cite{shani1984splines,
  chirikjian2013framed}) can be used as well.

\begin{figure}[htbp]
  \centering
    \includegraphics[scale=0.4]{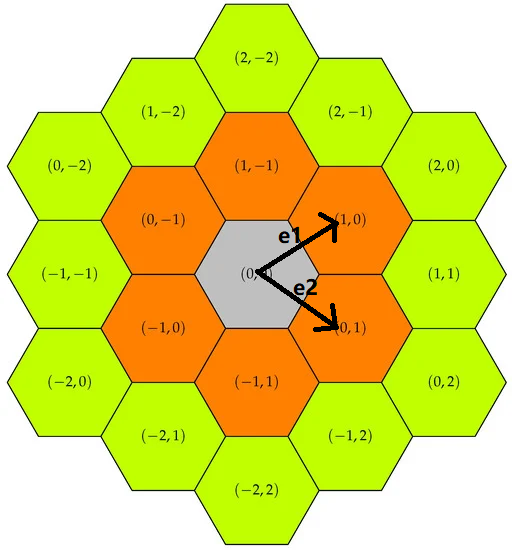}
  \caption{Section area view of the lane arrangement}
  \label{fig:lane-arrangement}
\end{figure}

Given a center curve in a lane, its parallel lanes are arranged in a
regular honeycomb (or hexagonal lattice) structure in the moving frame
(i.e.\ in the normal plane along the central curve), as shown in
Figure~\ref{fig:lane-arrangement}. This figure shows the central lane
in the grey hexagon, the other lanes are shown in orange and green. If
in this normal plane we define the point belonging to the center lane
as the origin and the two orthogonal unit vectors \vect{u_1} and
\vect{u_2} spanning the plane, then the mid-points of the parallel
lanes are arranged on a lattice with the two basis vectors
\begin{displaymath}
 \vect{e_1} = 2 \cdot r \cdot \vect{u_1}
\end{displaymath}
and
\begin{displaymath}
  \vect{e_2} = 2 \cdot r \cdot \cos\left(\frac{\pi}{3}\right) \cdot \vect{u_1}
  + 2 \cdot r \cdot \sin\left(\frac{\pi}{3}\right) \cdot \vect{u_2} 
\end{displaymath}
where $r$ is the radius of a lane (compare
Figure~\ref{fig:lane-arrangement}). Note that
$\norm{\vect{e_1}}=\norm{\vect{e_2}}=r$. Given these two vectors
spanning the lattice, the positions of all other lattice points can be
expressed (in the normal plane) as a linear combination of the two
vectors $\vect{e_1}$ and $\vect{e_2}$ with integer
coefficients. Hence, their positions are defined by a pair of
identifiers $(i, j)$, with the pair $(0,0)$ representing the central
lane of the road.  With this setup, drones flying in adjacent lanes
will remain at least $2 \times r$ meters apart.

\subsubsection{Chained Curves}

As stated in the overview, a lane can be made up of a number of curves
chained together, and here we state some additional assumptions and
conditions we use to facilitate this. A lane is made up of a finite
number of curves \fnct{\gamma_1}, \fnct{\gamma_2}, \ldots,
\fnct{\gamma_N} which are chained together (sufficiently) smoothly, by
which we mean that the starting point of curve \fnct{\gamma_2}
coincides with the end point of curve \fnct{\gamma_1} with at least
identical tangent vectors and second derivatives, that similarly the
starting point of \fnct{\gamma_3} coincides smoothly with the end
point of curve \fnct{\gamma_2} and so on.\footnote{Note that we assume
  that the parallel lanes are constructed from curves parallel to the
  given curves \fnct{\gamma_1}, \fnct{\gamma_2}, \ldots,
  \fnct{\gamma_N} of the central lane, so they have an equivalent
  sub-division into lanes.} Recall that we stipulate that our curves
use arc-length presentation, which normally implies for a curve
\fnct{\gamma_} that its parameter representation runs from zero to
$L(\gamma_i)$.  For the chained curves making up a lane, it is more
convenient to use the following parametrization:
\begin{itemize}
\item The first curve \fnct{\gamma_1} runs from zero to
  $L(\gamma_1)$
\item The second curve \fnct{\gamma_2} runs from $L(\gamma_1)$ to
  $L(\gamma_1) + L(\gamma_2)$
\item The third curve \fnct{\gamma_3} runs from
  $L(\gamma_1)+L(\gamma_2)$ to
  $L(\gamma_1) + L(\gamma_2) + L(\gamma_3)$, and so on.
\item Generally, the $i$-th curve runs from
  \begin{displaymath}
    a_i = \sum_{k=1}^{i-1} L(\gamma_k)
  \end{displaymath}
  to
  \begin{displaymath}
    b_i = \sum_{k=1}^i L(\gamma_k) = a_i + L(\gamma_i)
  \end{displaymath}
\end{itemize}
This parameterization has the advantage that the parameter values
actually measure how far we are across the entire lane. We refer to
this as the \textbf{chained arc-length representation} of a lane. This
is the parameterization we adopt for the remainder of the paper. We
will also refer to some parameter value $t$ as the parameter on the
current lane, and from using the chained arc-length representation it
is then straightforward to work out to which particular curve the
parameter $t$ belongs, namely the unique curve
$\fnct{\gamma_i} : [a_i,b_i]\mapsto\reals^3$ for which
$a_i \le t \le b_i$. holds.

\subsubsection{Distance Calculation and Parameter Conversion}
\label{subsubsec:drs:distance-calculation}

To judge the whether switching lane currently is safe, drones are
expected to learn the distances to neighbored drones on the same lane
and other drones on a target lane. In both cases it is not the
standard Euclidean distance that is relevant, but rather the
distance-along-the-lane or -curve. When a drone at parameter position
$s$ on a lane considers its distance to another drone at parameter
position $s'$ on the same lane, by the properties of arc-length
parameter representation their distance-along-the-lane is simply
$|s-s'|$.

Now suppose a drone is currently at position $s$ on a curve
$\fnct{\gamma}:[a,b]\mapsto\reals^3$ on its current lane, and switches
to a corresponding parallel curve
$\fnct{\gamma'}:[a',b']\mapsto\reals^3$ on an immediately neighbored
lane (in general the two curves will have a different length, and
therefore they will have different parameter intervals). In this
situation we assume that lane switching in the normal plane happens
instantaneously.  To calculate the resulting position parameter $s'$
on curve \fnct{\gamma'}, we use the following expression:
\begin{equation}
  \label{eq:parameter-conversion-parallel-curves-chained}
  s' = L(\gamma') \cdot \frac{s-a}{L(\gamma)} + a'
\end{equation}
and we use the shorthand $s' = \parconv{\gamma'}{\gamma}{s}$ to refer
to this conversion.

When a drone $D$ at position $s$ on curve \fnct{\gamma} on its current lane
wants to perform a parallel switch to a neighbored lane with current
curve \fnct{\gamma'} and wants to check for safety reasons the
distance to another drone at position $s''$ on curve \fnct{\gamma'}, we
do so by translating $D$'s position $s$ from curve \fnct{\gamma} to
its resulting position $s'$ on \fnct{\gamma'} (by evaluating $s' =
\parconv{\gamma'}{\gamma}{s}$) and then calculating the
distance-along-the-curve on \fnct{\gamma'}, i.e.
\begin{equation}
\label{eq:system-model:distance:distance-calculation-onThirdLane}
  |s'' - s'| = |s'' - \parconv{\gamma'}{\gamma}{s}|.
\end{equation}

\subsection{The \drselem{Curve} and \drselem{ChainedCurve} Elements}

In our XML-based representation we want to be able to distinguish
between the general concept of a curve and its attributes, and some
particular types of curves like for example Bezier curves, spline
curves or others, such that all the specific types have the attributes
of the general curve type and specific attributes as needed (e.g.\
control points for Bezier curves). To express this, XML has an
inheritance mechanism which we employ. In particular, we define a
general \drselem{Curve} element that we use to define further element
types. In our simulations we use a specialized type of curve, Bezier
curves, for which we additionally allow the specification of control
points (\drsattr{ControlPoint} attribute).

The \drselem{Curve} element representing a curve
$\fnct{\gamma}:[a,b]\mapsto\reals^3$ has the following attributes:
\begin{itemize}
\item \drsattr{Identifier}: A unique global identifier 
\item \drsattr{Description} (optional): A textual description
\item \drsattr{StartParameter}: The starting curve parameter value
  $a$ (in chained arc-length representation)
\item \drsattr{EndParameter}: The ending curve parameter value $b$ (in
  chained arc-length representation)
\item \drsattr{StartPoint}: The starting 3D point of the curve ($\gamma(a)$).
\item \drsattr{EndPoint}: The final 3D point of the curve ($\gamma(b)$).
\end{itemize}
The travel direction of the curve is given implicitly, drones always
move from the \drsattr{StartPoint} to the \drsattr{EndPoint}.

It is practically useful to have the ability to chain together several
curves, to allow modular construction of a longer lane from shorter,
given curve elements. This is facilitated by the
\drsattr{ChainedCurve} element, which has the following attributes and
sub-elements:
\begin{itemize}
\item \drsattr{Identifier}: A unique global identifier for the chained
  curve.
\item \drsattr{Description} (optional): A textual description
\item A list of one or more \drselem{Curve} elements, all smoothly
  connected and using chained arc-length representation.
\item \drsattr{StartParameter}: The starting chained curve parameter
  value (identical to \drsattr{StartParameter} attribute of the first
  \drselem{Curve} element).
\item \drsattr{EndParameter}: The final chained curve parameter
  value (identical to \drsattr{EndParameter} attribute of the last
  \drselem{Curve} element).
\end{itemize}
The travel direction of the \drselem{ChainedCurve} element is given by
the common travel direction of all its \drselem{Curve} sub-elements.

\subsection{The \drselem{Lane} Element}

The \drselem{Lane} element represents an individual lane belonging to
exactly one road. This lane includes a \drselem{ChainedCurve} element,
which represents the (chained) curve that drones flying on this lane
should track -- it is the center point of the lane. The travel
direction of the \drselem{Lane} element is the travel direction of the
\drselem{ChainedCurve} element it contains. The \drselem{Lane} has the
following attributes:
\begin{itemize}
\item \drsattr{Identifier}: A unique global identifier for the lane.
\item \drsattr{Description} (optional): A textual description.
\item One \drselem{ChainedCurve} sub-element, representing the
  directed flight path that drones on this lane have to take.
\item \drsattr{LaneIdentifier}: pair of integers $(i,j)$ identifying
  the lane (see Section~\ref{subsubsec:drs:parallel-curves}). In this,
  the lane identifier $(0,0)$ refers to the center lane of a road.
\item \drsattr{RoadIdentifier} (optional): the unique
  \drsattr{Identifier} of the \drselem{Road} element this lane belongs
  to.
\item \drsattr{ClosedCurves}: a list of integer curve indices
  (with reference to the contained \drselem{ChainedCurve}
  sub-element), specifying which of the curves are closed (i.e.\
  drones are not permitted to fly on the curve). Curves not referenced
  in this list are by definition open.
\end{itemize}
A lane can have further per-curve attributes not relevant here. It
should be noted that for a parallel lane (having
\drsattr{LaneIdentifier} different from $(0,0)$) the \drselem{Curve}
elements in the chained element are curves that are parallel to the
corresponding curve of the central lane (this needs to be guaranteed
by the design tool), and therefore it is possible to directly use the
parameter conversion method explained in
Section~\ref{subsubsec:drs:distance-calculation}.

\subsection{The \drselem{Road} Element}

A \drselem{Road} element holds one or more \drselem{Lane}
elements. One of these lanes is distinguished, we refer to it as the
\define{center lane} and it has \drsattr{LaneIdentifier} $(0,0)$. The
other lanes are parallel to the center lane. A \drselem{Road} element
has the following attributes and sub-elements:
\begin{itemize}
\item \drsattr{Identifier}: A unique global identifier for the road.
\item \drsattr{Description} (optional): A textual description.
\item A \drselem{Lane} sub-element representing the center lane.
\item One or more further \drselem{Lane} elements representing
  distinct lanes (with distinct \drsattr{LaneIdentifier} attributes)
  parallel to the center lane.
\item \drsattr{SpeedLimit}: this is the maximum speed that drones are
  allowed to have on this road.
\end{itemize}
All the lanes of the road have the same direction, and therefore the
road element as a whole also runs in one direction.

Note that for the purposes of this paper we have made the
\drsattr{SpeedLimit} a per-road attribute. Other choices are possible
as well, for example allowing for separate speed limit on the level of
individual lanes or even individual curves in a chain.

\subsection{The \drselem{Ramp} Element}

Ramps are specialized roads that include a single "always-open" lane,
which are used to provide additional entries and exits to the system
or to connect different roads (connecting a single lane from one road
to a single lane of another road).  Similar to roads, ramps are also
uni-directional. A valid ramp connects to at least one road. Depending
on the roads it connects to and its direction, a ramp can be
classified into three distinct categories:
\begin{itemize}
\item \textbf{On-ramp / Off-ramp}: To enable drones to enter or leave
  a road outside of its endpoints, we use on-ramps and
  off-ramps. These connect to exactly one lane on one end of the ramp,
  while the other end of the ramp serves as an access point for drones
  to enter or exit the road system. The classification of ramps
  depends on the type of connection; a "from-ramp-to-road" connection
  indicates an on-ramp, while a "from-road-to-ramp" connection
  indicates an off-ramp.
\item \textbf{Connecting-ramp}: Since roads are not directly connected
  to one another, we use connecting ramps to link two different roads,
  enabling smooth transitions and facilitating efficient
  road-switching. A connecting ramp between two roads is attached to
  one specific lane in either road to connect them, preserving
  orientation and allowing for smooth transition of a drone from the
  ramp onto the target lane (or vice versa).  Consequently, the route
  for a drone moving from Road A to Road B is "Road A → connecting
  ramp A and B → Road B."  This specific design ensures that drones
  remain confined to a single road and that each road operates as a
  one-way system.
\end{itemize}
The \drselem{Ramp} element has the following attributes and sub-elements:
\begin{itemize}
\item \drsattr{Identifier}: A unique global identifier for the ramp.
\item \drsattr{Description} (optional): A textual description.
\item A \drselem{Road} sub-element with exactly one lane that is open
  everywhere.
\item \drsattr{EntryRoadIdentifier} (optional): the unique
  \drsattr{Identifier} of the road the entry of the ramp is connected
  to. Drones can move from the road onto the ramp. If no
  \drsattr{EntryRoadIdentifier} is given, then the ramp is an on-ramp.
\item \drsattr{ExitRoadIdentifier} (optional): the unique
  \drsattr{Identifier} of the road the exit of the ramp is connected
  to. Drones can move from the ramp onto the road. If no
  \drsattr{ExitRoadIdentifier} is given, then the ramp is an
  off-ramp. If both \drsattr{EntryRoadIdentifier} and
  \drsattr{ExitRoadIdentifier} are given, then the ramp is a
  connecting ramp.
\item \drsattr{EntryLaneIdentifier} (optional): the lane identifier
  $(i,j)$ on the entry road the ramp is connected to (if any). This
  attribute must be defined when \drsattr{EntryRoadIdentifier} is
  defined.
\item \drsattr{EntryLaneParameter} (optional): the lane position
  parameter on the entry road where the ramp is attached to the entry
  lane. This attribute
  must be defined when \drsattr{EntryRoadIdentifier} is defined.
\item \drsattr{ExitLaneIdentifier} and \drsattr{ExitLaneParameter} are
  similar to \drsattr{EntryLaneIdentifier} and
  \drsattr{EntryLaneParameter} but in case an
  \drsattr{ExitRoadIdentifier} is set.
\end{itemize}

\subsection{The \drselem{DroneRoadSystem} Element}

The \drselem{DroneRoadSystem} element is the mandatory top-level
element. It contains at least one road and an arbitrary number of
ramps as possible sub-elements. It has at least the following
attributes and sub-elements:
\begin{itemize}
\item \drsattr{Name}: A human-readable name or description of the drone
  road system.
\item \drsattr{Version}: A version number for the drone road system.
\end{itemize}
There can be further attributes like a checksum (to verify integrity
of a drone road system description), or a global speed limit.




\section{System Model}
\label{sec:system model}

Before introducing our guidance algorithm, we first clarify our system
model, covering networking and physical layer, beacon contents, and
neighbor table.

\subsection{Networking and Channel}
\label{sec:system model:network}

To transmit the required safety data to neighbored drones, we adopt
the approach taken in vehicular networks
\cite{karagiannis2011vehicular}.  We assume all drones are equipped
with a WiFi adapter and configure it according to the (former) 802.11p
amendment used in the WAVE vehicular standards. Drones build an ad-hoc
network to establish direct communication between each other without
reliance on fixed infrastructure. The fixed physical-layer settings
are listed in Table \ref{tab:network setting}.

\begin{table}[!htbp]
    \caption{WiFi network setting}
  \begin{center}
    \begin{tabular}{|l|l|}
      \hline
      \textbf{Parameter} & \textbf{Value}                 \\
      \hline
      Center Frequency   & 2.437 GHz                   \\
      \hline
      Bandwidth          & 10 MHz                          \\
      \hline
      Modulation         & OFDM with QPSK               \\
      \hline
      Encoder            & rate 1/2 convolutional encoder \\
      \hline
      Data Rate          & 12 Mbps                         \\
      \hline
    \end{tabular}
    \label{tab:network setting}
  \end{center}
\end{table}

In the MAC layer, we employ the IEEE 802.11 EDCA access function which
is the default method for vehicular ad-hoc networks \cite{7093187,
  hasrouny2017vanet}. We use best effort (AC\_BE) as the access
category, setting the minimum and maximum contention window sizes to
31 and 1023, respectively, with an AIFS setting of 7. These settings
are conservative and can accommodate higher vehicle densities.

As a further assumption, periodic beacons are the only packet
type. These are broadcast locally and without acknowledgments, and no
other unicast or multicast traffic is present in the system, nor is
any external interference. Beacons may still experience delays (e.g.\
access delay) or packet loss from either direct or hidden-terminal
packet collisions. This kind of interference is for fixed beacon
length mainly controlled by parameters: beacon rate and transmit
power, and we will vary these parameters to explore their effects in
our simulations.

For the wireless channel model we use a simple log-distance path loss
model with given path loss exponent \cite{theodore2002wireless}. There
is no external interference and we also assume there to be no
multipath fading. We do not model shadowing from buildings or other
structures.

\subsection{Beacon Contents}
\label{sec:system model:beacon}

\begin{figure}[htbp]
  \centering
  \includegraphics[scale=0.5]{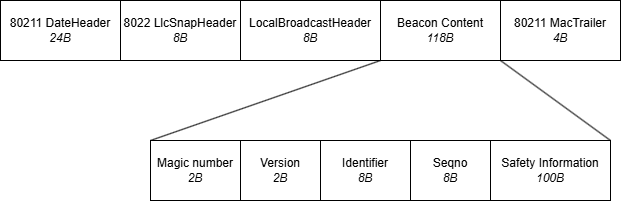}
  \caption{Beacon Content}
  \label{fig:beacon content}
\end{figure}

The beacon is the only packet type transmitted between drones, and in
this paper we assume that besides a unique identifier for a drone only
its position and velocity information in its data frame to conform to
the 'coordination-free' design principle. The detailed content
structure is shown in Figure \ref{fig:beacon content}.

The safety information block includes:
\begin{itemize}
\item \textbf{Position vector} $\vect{p}$ in $(x,y,z)$ form
\item \textbf{Velocity vector} $\vect{v}$ in $(x,y,z)$ form
\item \textbf{Road identifier} $I_r$
\item \textbf{Lane identifier} $(i,j)$
\item \textbf{Lane position parameter} $t$ which is between 0 and
  the length of the lane.
\end{itemize}

To allow us to focus purely on the effect of packet collisions on the
drone collision rate, we exclude other sources of positional
uncertainty. Specifically, we assume that the position and velocity
vectors $\vect{p}$ and $\vect{v}$ are always accurately measured by
drones using their GPS and accelerometers. These two parameters
provide the "absolute" position and velocity information within our
drone road system. However, since a single road in 3D space can
contain multiple curved lanes, relying on these coordinates to infer
road and lane information is not convenient. Therefore, we furthermore
assume that drones are aware of their current road ID, lane ID, and
relative position within their lane. The method for mapping \vect{p}
and \vect{v} onto road ID and lane ID is out of the scope of this
paper.  With these three additional parameters, drones can efficiently
compute their distance from others and the distance from the next
connecting ramp.

\subsection{Neighbor Table}
\label{sec:system model:neighbor}

Drones continuously receive beacons from others to stay aware of their
surroundings. When a beacon is received, the drone records the
information in its neighbor table for further decision-making. This
table is crucial for maintaining coordination-free guidance in the
drone road system, ensuring both safety and efficiency in complex
environments. In addition to the original data (listed in Section
\ref{fig:beacon content}) contained in the packet, the neighbor table
also includes the following:
\begin{itemize}
\item \textbf{Sender ID}: This is a unique identifier representing the
  sender, typically the MAC address.
\item \textbf{Sequence number}: Extracted from the beacon packet, and
  can be used to calculate packet loss rates.
\item \textbf{TimeStamp}: The last time a position update was received
  from the sender. This allows to check how old the information is,
  and also enables scrubbing, i.e.\ pushing out information from the
  neighbor table that is too old.
\item \textbf{Distance to sender} $d_{to}$: This is the
  distance-along-the-curve used when the current drone intends to
  switch to the sender's lane, calculated according to
  Equation~\eqref{eq:parameter-conversion-parallel-curves-chained}. If
  the sender is on a different road or ramp but remains directly
  connected with the current road or ramp, the value should be the sum
  of two parts: the distance from one drone to the connection point
  and the distance from the connection point to the other drone.
\item \textbf{Distance from sender} $d_{from}$: Similar to $d_{to}$,
  but it is used when the sender intends to switch to the current
  drone's lane. 
\end{itemize}
The neighbour table can have further entries, for example an entry for
the estimated time until collision. These entries do not play a role
for our guidance algorithm.

\section{The Short-Term Decentralized Greedy (STDG) Guidance Algorithm}
\label{sec:algorithm}

The short-term decentralized greedy (STDG) guidance algorithm for the
drone road system is introduced here. This algorithm evolved from the
greedy lane-switching algorithm described in
\cite{qu2022sensorless}. The upgrade expands its application from a
single straight road segment to a comprehensive road system that
includes curves and interconnections between roads.

\subsection{Preliminaries}
\label{sec:algorithm:preliminaries}

STDG aims to make the short-term decisions for route guidance and
collision avoidance, assuming that the end-to-end path (given as a
sequence of roads and connecting ramps) is known. The precise
description of this end-to-end path and the method of its computation
are outside the scope of this paper, but our guidance algorithm needs
information about the position of the next connecting ramp, so that it
can choose its lanes so as to be able to reach the connecting ramp
once it gets closer. This next-ramp information is continuously
updated in the background. We further assume that all drones are
familiar with the entire drone road system and that, apart from other
drones in motion, no other obstacles or buildings intersect with the
road system.

On the highest level, the algorithm will have to make a decision about
the speed of the drone and the lane to use for the next few seconds
(chosen from a suitable subset of its immediate neighbor lanes, the
set of \define{candidate lanes}), until the next invocation of the
algorithm. For each candidate lane $(i',j')$ we determine the possible
speed $v'$ we can use on that lane. We evaluate a cost function
$c((i',j'),v')$ for each candidate lane and its achievable speed, and
then select the candidate lane / speed combination $((i^*,j^*),v^*)$
which minimizes the costs (with ties being broken arbitrarily).

The key part of the algorithm then is clearly the design of the cost
function \fnct{c}, which can account for a range of factors:
\begin{itemize}
\item The achievable speed $v'$ on a candidate lane and its
  relationship to the preferred speed $\vpref$ of the drone.
\item The lane position: the drone will move along roads and will
  occasionally have to switch roads through specific connecting ramps,
  as prescribed by the end-to-end path. As long as a drone is far away
  from the next connecting ramp it needs to use (which is attached to
  a particular lane on the current road -- the \textbf{target lane} --
  at a known parameter point), the choice of the next lane to use does
  not matter much, but as we get closer to the next connecting point,
  we want to be close to or even on the target lane.
\item Other drones: the speed and current lanes of other drones, which
  will be key in assessing collision risk.
\item Parallel collision risk: in a simple example of a parallel
  collision two drones on three lanes are involved. The three lanes
  are arranged in a line, and the two drones are on either of the
  outer lanes while the middle lane is free. If both drones are more
  or less parallel to each other and decide to switch to the middle
  lane at the same time (which they both assess as free), then a
  collision can ensue.
\end{itemize}

\subsection{Notations}
\label{sec:algorithm:notations}

We define the followings notations which will be used in the remainder
of the algorithm description:
\begin{itemize}
\item The current lane is lane $(i,j)$.
\item The set $\neighlanes{i}{j}$ of immediate neighbor lanes of lane
  $(i,j)$ includes those existing lanes which can be reached with one
  ``lane hop'' from lane $(i,j)$. Note that
  $(i,j)\not\in\neighlanes{i}{j}$.
\item The set $\candlanes{i}{j} = \neighlanes{i}{j} \cup \set{(i,j)}$
  is the set of candidate lanes that the current drone can switch to
  from lane $(i,j)$ (including staying on the same lane). In this
  paper we stipulate that the set of candidate lanes
  $\candlanes{i}{j}$ does not include lanes that will close soon
  (except when being identical to the target lane and the target
  position being closer than the next closing
  position). Alternatively, they could be included but marked with a
  high position cost.
\item For a given lane $(i,j)$ we denote by \drones{i}{j} the current
  set of drones on lane $(i,j)$ -- at least the ones that the current
  drone is aware of (i.e.\ has stored in its neighbor table). In some
  places we work with a set $\mathcal{L}$ of lanes. In such a case,
  the notation \dronesunion{\mathcal{L}} refers to the union of all
  the sets \drones{i}{j} for all $(i,j)\in\mathcal{L}$.
\item For two parallel lanes $(i,j)$ and $(i',j')$ on the same road
  the expression $\hopdistance{(i,j)}{(i',j')}$ denotes the 'hop
  distance' of these two lanes, i.e.\ the smallest length of all the
  chains $(i_1,j_2), (i_2,j_2), \ldots, (i_K,j_K)$ of successively
  immediately neighboured lanes with $(i_1,j_1)=(i,j)$ and
  $(i_K,j_K)=(i',j')$. We use the following expression for the hop
  distance (detailed explanation is in Appendix \ref{sec:app:explanation-hop-distance}):
  \begin{equation}
    \hopdistance{(i,j)}{(i',j)} = \left\{\begin{array}{r@{\quad:\quad}l}
      |i-i'| + |j-j'| & (i-i')\cdot(j-j') \ge 0 \\
      \max\set{|i-i'|, |j-j'|} & \mbox{otherwise}
    \end{array}
    \right.
  \end{equation}
\item For lane $(i,j)$, \closingpoints{i}{j} is the set of
  all chained arc-length parameter points where the lane closes
  (i.e.\ where a closed component curve starts).
\item For lane $(i,j)$, \ramppoints{i}{j} is the set of all arc-length
  parameter points where the lane is attached to any type of outgoing
  ramp (off-ramp or outgoing connecting ramp).
\item For lane $(i,j)$, \dronepoints{i}{j} is the set of all
  arc-length parameter points of drones on that lane (according to the
  neighbor table).
\item For a given set $T=\set{t_1, t_2, \ldots, t_K}$ of chained
  arc-length parameter points of a lane, denote by \aheadpoints{T}{t}
  the subset of all $t_i \in T$ such that $t_i > t$, i.e.\ the set of
  all points of the given set that still lie ahead of point
  $t$. Denote by \behindpoints{T}{t} the subset of all $t_i \in T$
  such that $t_i \le t$, i.e.\ the set of all points of the given set
  that are behind of point $t$.
\item Building on this, for a given set $T=\set{t_1, t_2, \ldots,
    t_K}$ of parameter points of a lane, denote by
  \begin{equation}
    \closestaheaddist{T}{t} = \left\{\begin{array}{r@{\quad:\quad}l}
      \min \setof{t' - t}{t' \in \aheadpoints{T}{t}} & \aheadpoints{T}{t}\ne\emptyset\\
      \infty & \mbox{otherwise}
    \end{array}
    \right.
  \end{equation} the distance from the current position $t$ to the
  smallest parameter point $t'\in T$ ahead (if exists).
\item Similarly, denote by
  \begin{equation}
    \closestbehinddist{T}{t} = \left\{\begin{array}{r@{\quad:\quad}l}
      \min \setof{t - t'}{t' \in \behindpoints{T}{t}} & \behindpoints{T}{t}\ne\emptyset\\
      \infty & \mbox{otherwise}
    \end{array}
    \right.
  \end{equation}
  the distance from the current position $t$ to the largest parameter
  point $t'\in T$ behind (if exists). 
\item For a drone $A$, \dlane{A} is its current lane. We do not
  include its current road in our notation, we assume that to be
  understood.
\item For a drone $A$, \dpos{A} is its current position on the current
  lane (i.e.\ its parameter in the chained arc-length
  parameterization).
\item For a drone $A$, \dspeed{A} is its current speed on the current
  lane.
\item A high cost value $C_{max}$ which is used for marking the
      candidate lanes are not feasible for lane switching in some
      situations.
      
\end{itemize}

\subsection{Algorithm Inputs}
\label{sec:algorithm:inputs}

These inputs are re-calculated at the start of every invocation of our
algorithm based on the current position of the drone (current road,
lane and chained arc-length parameter), the knowledge of the drone
road system map, and the neighbour data the drone has received from
other drones and stored in the neighbour table. In the following, we
use letters $s$ and $t$ for parameter points (with indices, as
needed), and these always refer to chained arc-length parameters on
their respective lanes.
\begin{itemize}
\item The current lane $(i,j)$ and parameter point $t_{i,j}$ on that
  lane (which, as a reminder, represents the distance-along-the-curve
  between the start of lane $(i,j)$ and the current point on the
  lane), as well as the current speed $v$.
\item The target lane $(i_T,j_T)$ (i.e.\ the lane we have to be on the
  next time we need to use a connecting ramp to switch roads, if any,
  or an off-ramp to leave the drone road system) and associated
  parameter point $t_{i_T,j_T}$ on that lane specifying the point
  (called the \textbf{target point}) where we will have to be on that
  lane in order to be able to take the target ramp. Furthermore, we
  are given the distance $d_T$ we would have to travel along the
  target lane to reach the target point, pretending that we currently
  \emph{are} on the target lane (with current parameter value
  $\parconv{\gamma_{i_T,j_T}}{\gamma_{i,j}}{t_{i,j}}$ on the target
  lane). This distance is given by
  \begin{equation}
    \label{eq:target-point-distance}
    d_T = t_{i_T,j_T} - \parconv{\gamma_{i_T,j_T}}{\gamma_{i,j}}{t_{i,j}}
  \end{equation}
\item The current set $\candlanes{i}{j}$ of candidate
  lanes as explained above. Note that a lane $(i',j')$ is only
  considered as a candidate lane when the distance towards the next
  closing point is larger than a given threshold $\epsilon_0 >
  0$, i.e.\ when
  \begin{equation}
    \closestaheaddist{\closingpoints{i'}{j'}}{\parconv{\gamma_{i',j'}}{\gamma_{i,j}}{t_{i,j}}}
    \ge \epsilon_0
  \end{equation} holds. For each candidate lane
  $(i',j')\in\candlanes{i}{j}$ we additionally collect the following
  information:
  \begin{itemize}
  \item The distance-along-the-lane \distahead{i'}{j'} to the closest
    drone ahead on lane $(i',j')$. This is defined as
    \begin{equation}
      \distahead{i'}{j'}
       =  \closestaheaddist{\dronepoints{i'}{j'}}{\parconv{\gamma_{i',j'}}{\gamma_{i,j}}{t_{i,j}}}\\
    \end{equation}
    In calculating this distance-along-the-line to the ahead drone, we
    pretend again that the current drone is actually positioned on
    lane $(i',j')$ and has position / parameter
    $\parconv{\gamma_{i',j'}}{\gamma_{i,j}}{t_{i,j}}$ on that lane.
    \item To assess the risk of a parallel collision happening if we
    were to switch from lane $(i,j)$ to our candidate lane $(i',j')
    \ne (i,j)$, we count the number of drones which pose potential
    collision risks. We separate this into two different numbers: the
    number $\numnearbysame{i'}{j'}$ of nearby drones on the candidate
    lane $(i',j')\ne(i,j)$ itself, and the number
    $\numnearbyneigh{i'}{j'}$ of nearby drones located on neighbor
    lanes of $(i',j')$ (including $(i,j)$). The sets of these drones
    are defined as:
    \begin{eqnarray}
      \Dronesnearbysame{i'}{j'} & = & \setof{A' \in \dronesunion{\set{(i',j')}}}
                                  {|\parconv{\gamma_{i',j'}}{\gamma_{i,j}}{t_{i,j}} - \dpos{A'}| < \epsilon_1}\\
      \Dronesnearbyneigh{i'}{j'} & = &  \setof
                            {A'' \in \dronesunion{\neighlanes{i'}{j'}}}
                            {|\parconv{\gamma_{i',j'}}{\gamma_{i,j}}{t_{i,j}} -
                            \parconv{\gamma_{i',j'}}{\gamma_{i'',j''}}{\dpos{A''}}|
                            < 2 \epsilon_1}
    \end{eqnarray} where $\epsilon_1 > 0$ is a threshold parameter,
    and the choice of $2\epsilon_1$ in the second condition is made to
    provide some safety margin due to uncertainties in node
    positions. The numbers of nearby drones on these two sets are then
    given by
    \begin{eqnarray}
      \label{eq:algorithm:numnearbysame}
      \numnearbysame{i'}{j'} & = & \magnitude{\Dronesnearbysame{i'}{j'}}\\
      \label{eq:algorithm:numnearbyneigh}
      \numnearbyneigh{i'}{j'} & = & \magnitude{\Dronesnearbyneigh{i'}{j'}}
    \end{eqnarray}
  \item The speed $v_{i',j'}$ of the closest drone ahead on lane
    $(i',j')$. As before, if there is no ahead drone then we set
    $v_{i',j'}=\infty$.
  \item We determine the achievable speed $v^*_{i',j'}$ on candidate
    lane $(i',j')$ as
    \begin{equation}
      \label{eq:candidate-achievable-speed}
      v^*_{i',j'} = \left\{\begin{array}{r@{\quad:\quad}l}
        \min\set{\vpref,\vmax, d_{target}/s} & \distahead{i'}{j'} > \epsilon_2 \\
        \min\set{\vpref, v_{i',j'},\vmax, d_{target}/s} & \distahead{i'}{j'} \le \epsilon_2 \\
      \end{array}
      \right.
    \end{equation} where $\vmax$ is the maximum speed allowed on the
    current lane or road.  This choice reflects our assumption that if
    the closest-ahead drone on lane $(i',j')$ has a sufficiently large
    distance-along-the-lane from us, we can proceed (at least for a
    short while) at our preferred speed, otherwise we adopt the speed
    of the ahead drone, our preferred speed, or $\vmax$, whichever is
    smallest. In all cases the speed is limited by the maximum allowed
    speed $\vmax$ on the current road or lane, and is also limited by the distance to its target point ($d_{target}$, if any) divided by the switching time $s$, to ensure that the drone does not miss the target point. The value $\epsilon_2>0$ is a parameter to be chosen.
  
  \end{itemize}
\item A set
  $\blockinglanes{i}{j} \subset \candlanes{i}{j}
  \setminus\set{i_T,j_T}$ of so-called non-target or blocking
  lanes. These blocking lanes include all candidate lanes distinct
  from the target lane $(i_T,j_T)$ which have a connecting ramp ahead
  and the distance to that out point is closer than a threshold value
  $\epsilon_0$. These lanes might attract \emph{other} drones which
  have to use it to switch roads. On such hotspots, these blocking
  lanes are preferably avoided due to potential build up of queues.
\end{itemize}

\subsection{Overall Cost Model}
\label{sec:algorithm:costModel}

To a given candidate lane $(i',j')\in\candlanes{i}{j}$ and its
associated achievable speed $v^*_{i',j'}$ (compare
Equation~\eqref{eq:candidate-achievable-speed}) we assign a cost value
as follows:
\begin{equation}
  c((i',j'), v^*_{i',j'}) =
  c_s(v^*_{i',j'})
  + \kappa_1 \cdot c_p(i',j')
  + \kappa_2 \cdot c_c((i',j'), v)
\end{equation}
where $\kappa_1, \kappa_2 \ge 0$ are non-negative weight parameters,
$\fnct{c_s}$ is the so-called speed cost function which assigns a cost
to the difference between the achievable speed on that lane and the
drones preferred speed, $\fnct{c_p}$ is the so-called position cost
which expresses how strongly we need to drift to our target lane as we
come closer to the next target point, and $\fnct{c_c}$ is the
so-called collision cost expressing how likely we might experience a
parallel collision. We discuss these individual cost functions in
turn.

\subsection{Speed Costs}
\label{sec:algorithm:speed}

For a given candidate lane $(i',j')\in\candlanes{i}{j}$ and candidate
speed value $v$ we calculate its speed cost as
\begin{equation}
  c_s(v) = \left(\vpref - v\right)^2
\end{equation}
This model punishes larger speed differences more strongly.

\subsection{Position Costs}
\label{sec:algorithm:position}

\begin{figure}[!htb]
  \centering
  \subfigure[With T ($(i_T,j_T) \ne \mbox{none}$) ]{
    \begin{tikzpicture}[scale=0.3]
      \begin{scope}[%
          every node/.style={anchor=west,
            regular polygon, 
            regular polygon sides=6,
            draw,minimum width=3cm,
            outer sep=0,},transform shape]
        \node[fill=orange] (A) {(1,-2)};
        \node[fill=OrangeRed] (B) at (A.corner 1) {(2,-2)};
        \node[fill=orange] (C) at (A.corner 5) {(1,-1)};
        \node[anchor=east,fill=Yellow] (D) at (A.corner 4) {(0,-2)};
        \node[fill=Yellow] (E) at (D.corner 5) {(0,-1)};
        \node[anchor=east,fill=lime] (F) at (E.corner 4) {T};
        \node[fill=red] (G) at (C.corner 1) {B};
        \node[fill=Yellow] (H) at (F.corner 5) {(-1,0)};
        \node[anchor=east,fill=Yellow] (I) at (H.corner 4) {(-2,0)};
        \node[fill=orange] (J) at (H.corner 1) {(0,0)};
        \node[fill=OrangeRed] (K) at (J.corner 1) {(1,0)};
        \node[fill=red] (L) at (K.corner 1) {(2,0)};
        \node[fill=orange] (M) at (I.corner 5) {(-2,1)};
        \node[fill=orange] (N) at (M.corner 1) {(-1,1)};
        \node[fill=OrangeRed] (O) at (N.corner 1) {(0,1)};
        \node[fill=red] (P) at (O.corner 1) {(1,1)};
        \node[fill=OrangeRed] (Q) at (M.corner 5) {(-2,2)};
        \node[fill=OrangeRed] (R) at (Q.corner 1) {(-1,2)};
        \node[fill=red] (S) at (R.corner 1) {(0,2)};
        draw [->] (J.center) -- (K.center);
      \end{scope}
    \end{tikzpicture}
  }\hspace{0.2cm}
  \subfigure[Without T ($(i_T,j_T) = \mbox{none}$)]{
    \begin{tikzpicture}[scale=0.3]
      \begin{scope}[%
          every node/.style={anchor=west,
            regular polygon, 
            regular polygon sides=6,
            draw,minimum width=3cm,
            outer sep=0,},transform shape]
        \node[fill=lime] (A) {(1,-2)};
        \node[fill=lime] (B) at (A.corner 1) {(2,-2)};
        \node[fill=lime] (C) at (A.corner 5) {(1,-1)};
        \node[anchor=east,fill=lime] (D) at (A.corner 4) {(0,-2)};
        \node[fill=lime] (E) at (D.corner 5) {(0,-1)};
        \node[anchor=east,fill=red] (F) at (E.corner 4) {B};
        \node[fill=red] (G) at (C.corner 1) {B};
        \node[fill=lime] (H) at (F.corner 5) {(-1,0)};
        \node[anchor=east,fill=lime] (I) at (H.corner 4) {(-2,0)};
        \node[fill=lime] (J) at (H.corner 1) {(0,0)};
        \node[fill=lime] (K) at (J.corner 1) {(1,0)};
        \node[fill=lime] (L) at (K.corner 1) {(2,0)};
        \node[fill=lime] (M) at (I.corner 5) {(-2,1)};
        \node[fill=lime] (N) at (M.corner 1) {(-1,1)};
        \node[fill=lime] (O) at (N.corner 1) {(0,1)};
        \node[fill=lime] (P) at (O.corner 1) {(1,1)};
        \node[fill=lime] (Q) at (M.corner 5) {(-2,2)};
        \node[fill=lime] (R) at (Q.corner 1) {(-1,2)};
        \node[fill=lime] (S) at (R.corner 1) {(0,2)};
        draw [->] (J.center) -- (K.center);
      \end{scope}
    \end{tikzpicture}
  }\hspace{0.2cm} \caption{Visualization of the position cost, lanes
  with the darker color have higher position cost, B means blocking
  lane, and T means target lane}
  \label{fig:chapter3:short_position}
\end{figure}
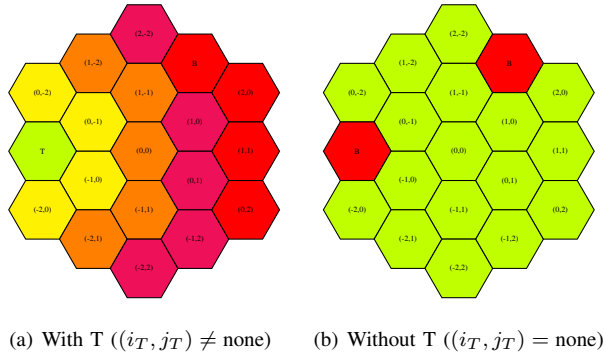

For associating a position cost $c_p(i',j')$ to candidate lane
$(i',j')$ we take into account the following factors:
\begin{itemize}
\item How close we are to the target point (if any) -- this distance
  is given by $d_T$ from Equation~\eqref{eq:target-point-distance} --
and the hop distance between our current lane and the target lane --
we would like to express that as long as the distance-along-the-lane
to the target point is large, the hop count does not matter a lot, but
as we get closer to the target point, it will matter more and more.
\item If the lane is a blocking lane, i.e.\
  $(i',j')\in\blockinglanes{i}{j}$, then we expect there to be
  potential for other drones queueing up there (at least if the
  closest ahead ramp is reasonably close) and we would prefer to avoid
  it.
\end{itemize}
With these considerations, the current prescription for the position
cost is:
\begin{equation}
  c_p(i',j') = \left\{\begin{array}{r@{\quad:\quad}l}
    f(d_T) \cdot \hopdistance{(i',j')}{(i_T,j_T)}
          & ((i_T,j_T) \ne \mbox{none}) \wedge((i',j') \notin \blockinglanes{i}{j}) \\
    C_{max} & ((i',j') \ne (i_T,j_T)) \wedge ((i',j') \in \blockinglanes{i}{j})\\
    0      & \mbox{otherwise}
  \end{array}
  \right.
\end{equation}
Figure \ref{fig:chapter3:short_position} shows the visualization of
the function.

In this expression, recall that \hopdistance{(i',j')}{(i_T,j_T)}
refers to the hop distance between two lanes, and the function
$f:[0,\infty)\mapsto[0,\infty]$ is a monotonically decreasing function
(which should have small values for large arguments, and large values
  for small arguments). In this paper we have chosen $f(d)$ to be
  the following function:
  \begin{equation}
    f(d) =  \left\{\begin{array}{r@{\quad:\quad}l}
     20 & 0\le d \le 15 \\ 
     -\frac{2}{3} d + 30 & 15 \le d \le 30 \\
     \frac{300}{d} & d \ge 30
    \end{array}
    \right.
    \label{eq:fd}
  \end{equation}


\subsection{Collision Costs}
\label{sec:algorithm:collision}

The collision cost $c_c((i',j'), v)$ for candidate lane $(i',j')$ and
current drone speed $v$ is designed to reflect the risk of a parallel
collision (in which the drone switching lanes collides with another
drone switching simultaneously to the same lane), and collisions with
drones already present on the target lane $(i,j)$. As a proxy to
assess these risks we use the numbers \numnearbysame{i'}{j'} and
\numnearbyneigh{i'}{j'} (compare
Equations~\eqref{eq:algorithm:numnearbysame} and
\eqref{eq:algorithm:numnearbyneigh}) of nearby drones for these two
different cases.

In the expression for the collision cost we propose below, we also
introduce the concept of priority of a drone: suppose the current
drone is on lane $(i,j)$ and considers neighbor lane $(i',j') \ne
(i,j)$ as the candidate lane. Even if the set
\Dronesnearbyneigh{i'}{j'} of nearby drones on lane $(i',j')$ is
non-empty, we can consider switching onto that lane if the current
drone is \emph{ahead} of all the nearby drones on the neighbor lanes
of $(i',j')$ and can choose a speed $v$ at least as large as the speed
of the closest drone on neighbor lanes of $(i',j')$ (but not exceeding
the preferred speed $\vpref$).

For the current drone A on lane $(i,j)$, the distance to closest
behind drone over all nearby drones \Dronesnearbyneigh{i'}{j'} is:
\begin{equation}
    d^{nearby}_{i',j'} = \min\setof{\parconv{\gamma_{i',j'}}{\gamma_{i,j}}{\dpos{A}} - \parconv{\gamma_{i',j'}}{\gamma_{\dlane{A'}}}{\dpos{A'}}}{A'\in\Dronesnearbyneigh{i'}{j'}}
\end{equation}
(or $d^{nearby}_{i',j'} = \infty$ when
$\Dronesnearbyneigh{i'}{j'}=\emptyset$), where calculating distance
between two drones on the third lane is carried out using
Equation~\eqref{eq:system-model:distance:distance-calculation-onThirdLane}
introduced in the Section
\ref{subsubsec:drs:distance-calculation}. Note that the value is
sign-sensitive, a positive value means that the other drone is behind
the current one, and vice versa.  Therefore, for a positive
$d^{nearby}_{i',j'}$ the distances to all nearby drones is positive,
meaning that the current drone A is in front of all nearby drones.
Negative values of $d^{nearby}_{i',j'}$ are not meaningful in this
context. If we name the closest behind drone (over
\Dronesnearbyneigh{i'}{j'}) of the current drone A as
$A^{nearby}_{i',j'}$, then the current drone $A$ has priority to
switch to lane $(i',j')$ when:
\begin{equation}
    \mbox{hasPriority} = \left\{\begin{array}{r@{\quad:\quad}l}

    \mbox{true} & d^{nearby}_{i',j'} > \epsilon_3\\

    \mbox{true} & (0 < d^{nearby}_{i',j'} \leq \epsilon_3)\wedge
    (v > \dspeed{A^{nearby}_{i',j'}})\\

    \mbox{false} & \mbox{otherwise} 
    
    \end{array}
    \right.
\end{equation}
where $\epsilon_3$ is a threshold parameter which should be positive
and smaller than $\epsilon_1$.

With this, the collision cost function is given by
\begin{equation}
  c_c((i',j'),v) = \left\{\begin{array}{r@{\quad:\quad}l}
    C_{max} & ((i',j')\ne(i,j)) \wedge (\numnearbysame{i'}{j'}>0) \\
    0 & ((i',j') \ne (i,j)) \wedge (\numnearbysame{i'}{j'}=0) \wedge \mbox{hasPriority} \\ 
    0 & (i',j') = (i,j) \\
    \numnearbyneigh{i'}{j'} & \mbox{otherwise}
  \end{array}
  \right.  
  \label{eq:collision_cost}
\end{equation}
  
\begin{figure}[htbp]
  \centering
  \includegraphics[scale=0.8]{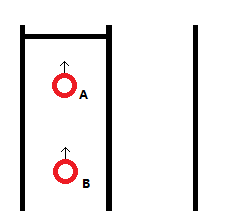}
  \caption{A situation where two drones are locked without setting the priority.}
  \label{fig:two_drones_locked}
\end{figure}

The priority term exempts a single drone, allowing it the maximum freedom to
switch lanes and thus preventing a scenario where all drones become
obstructed by others. An example is shown in Figure
\ref{fig:two_drones_locked}. The lane where drones A and B are
currently located will close soon, requiring both drones to switch to
the right neighbor lane. However, without a priority mechanism, the
drones will treat each other as a potential risk and apply the last
case of calculating $C_c$, resulting in the same cost values for their
decisions. As a result, both drones may come to a full stop
simultaneously, causing a deadlock situation. The priority mechanism
solves this issue by allowing drone A to set its collision cost to
zero, ensuring that drone A always chooses to switch lanes. This
prevents the drones from remaining stuck in their current lanes.

\subsection{Make a Decision}
\label{sec:algorithm:decision}

The cost model calculation returns a positive value representing the
weighted cost associated with each candidate lane and related
feasible speed $v_{i^\prime,j^\prime}^*$ (compare
Equation~\eqref{eq:candidate-achievable-speed}). We then select an
output pair $((i_o, j_o), v_{i_o,j_o})$ based on the following
criteria (in which we use the abbreviation $\tau = \setof{((i^\prime,
j^\prime),v_{i^\prime,j^\prime}^*)}{(i^\prime, j^\prime) \in
\candlanes{i}{j}}$):
\begin{equation}
  ((i_o, j_o), v_{i_o,j_o}) =
  \left\{\begin{array}{r@{\quad:\quad}l}
  \underset{((i^\prime, j^\prime), v_{i^\prime, j^\prime}^*) \in \tau}{\argmin} c(i^\prime, j^\prime) & \underset{((i^\prime, j^\prime), v_{i^\prime, j^\prime}^*) \in \tau}{\min}c(i^\prime, j^\prime) < C_{max} \\
  ((i, j), 0) & \underset{((i^\prime, j^\prime), v_{i^\prime, j^\prime}^*) \in \tau}{\min}c(i^\prime, j^\prime) \geq C_{max}
  \end{array}
  \right.  
\end{equation}
When the smallest cost over all candidate lanes is no smaller than
$C_{max}$, this indicates that all candidate lanes (including the
current lane) are infeasible for lane switching. In that case (or when
there is no candidate lane at all -- which can arise when the road
only has one lane which ends or closes soon), the drone will stop
(i.e.\ choose speed zero), and then wait for divine intervention or
depletion of its energy supply (whichever comes first). A special case
arises if the drone is currently on a connecting or on-ramp and
approaches the end of that ramp where it merges with a target lane on
another road. In this case, and if the current drone does not have any
other drone ahead of it on the ramp, the current drone also takes into account
the speed of drones on target lane to make decisions.

If multiple lanes have the same minimal cost value, we select the
current lane if it is among them; otherwise, we randomly choose one of
the other mimimal-cost lanes whose hop distance to the target lane is
smallest -- further ties are broken randomly. After obtaining the
output pair, it is necessary to determine whether the drone needs to
stop to ensure it does not miss the road switching point. The process
is:
\begin{equation}
  v_o =
  \left\{\begin{array}{r@{\quad:\quad}l}
  0 & (\parconv{\gamma_{i_T,j_T}}{\gamma_{i,j}}{t_{i,j}} - t_{i_T,j_T}) < \epsilon_0 \\
  v_{i_o,j_o} & \mbox{otherwise}
  \end{array}
  \right.    
\end{equation}
Then the final output is $((i_o,j_o), v_o)$, the drone should
immediately start switching to the lane $(i_o,j_o)$ with the speed
$v_o$.

\subsection{Step 4: Schedule the Next Run}
\label{sec:algorithm:s4}

The period of running algorithm $u$ depends on the current speed of
the drone. In the normal case (i.e.\ speed is not zero), and while the
current drone is flying within a lane, $u$ is set to a fixed value --
in our simulations we use 2 seconds. If the current drone currently
has zero speed, the algorithm is invoked more frequently to reduce the
average waiting time when escaping traffic congestion. In our
simulations we use a period of 0.5 seconds. The algorithm does not run
while the current drone is in the process of switching lanes.

When the drones finish their lane/road/ramp switching, they will
immediately start a new invocation of the algorithm to adjust their
speed based on the current situation. This could help prevent some
concurrent collisions (e.g., when two drones switch to the same lane
at a similar time, the rear drone can immediately detect the front
drone after switching and then adjust its speed or switch to another
lane).


\section{Simulation Results}
\label{sec:results}

In this section, we present simulation results to evaluate the
performance of the STDG algorithm. After explaining the simulation
setup and key metrics, we first investigate the impact of key MAC/PHY
parameters (beaconing rate and transmit power) and identify suitable
settings for our further investigations, in which we assess the impact
of the parameters of the STDG algorithm in a sensitivity
study. Finally, we discuss and evaluation the algorithm based on the study results and figure out some potential issues observed during the
simulations.

\subsection{Simulation environment and Assumptions}
\label{sec:results:environment-and-assumptions}

\begin{figure}[htbp]
  \centering
  \includegraphics[scale=0.4]{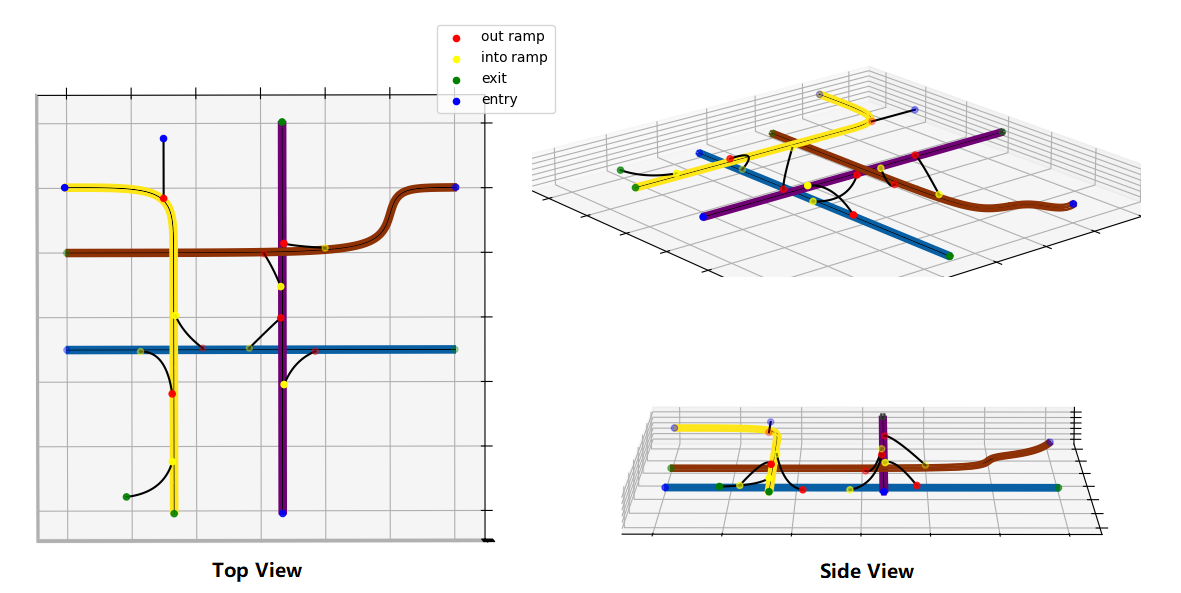}
  \caption{Drone Road System Used in the Simulations}
  \label{fig:drsInSim}
\end{figure}

To clarify the bounds of the simulation, we make the following
assumptions:
\begin{itemize}
\item We treat each drone as a point in the simulation environment, a
      collision happens when the distance between two drones is closer
      than the minimal safety distance.
\item Drones periodically perform the neighbor table cleaning up
      process, during which they remove outdated information after a
      configurable timeout.
\item After a collision, all involved drones immediately vanish from
      the simulation. This approach eliminates the possibility of
      chain collisions or secondary events caused by the initial
      collision (like a falling drone crashing into another
      one). However, other drones in the simulation will still store
      outdated information about the removed drones in their neighbour
      table until the next cleaning up process, which may impact
      decision-making.
\item The road system used in the simulation (shown in Figure
      \ref{fig:drsInSim}) resembles an intersection layout with four
      main roads extending in different directions and seven
      connecting lanes linking them. 
      
\item Drones are injected into the simulation with random starting and
      destination lanes or on/off-ramp. Due to this random assignment,
      over 80\,\% of drones are required to pass through a connecting
      lane to reach their destinations. In the given road system, this
      configuration can potentially result in significant traffic
      congestion around connecting lanes when drone density is high,
      as drones must wait their turn and queues can build up.
      
\item Drones are generated at time intervals that follow an
      exponential distribution, defined by a parameter called the
      \textbf{generation rate}, which represents how many drones will
      be generated on average per second per lane. Before the
      generated drones are injected into the simulation environment, a
      pre-collision-check process is performed to ensure that they do
      not collide with other existing drones at the time of
      injection. If the drone does not pass the check, we simply drop
      it. We refer to the actual rate at which drones are injected
      into the system as the \textbf{injection rate} which is less
      than or equal to the generation rate.
\item When drones reach their destination, they are removed from the
      simulation, but other nodes might still retain information about
      them in their neighbor table until the next clean up.
\item There is no external interference that could disrupt the
      beaconing communication between drones.
\end{itemize}

Our simulation \footnote{The source code is in \url{https://github.com/zqu14/STDG}} is built within the OMNeT++ discrete-event simulation
framework\footnote{\url{https://omnetpp.org/}} together with the INET
module library.\footnote{\url{https://inet.omnetpp.org/}} The basic
fixed simulation parameters are shown in
Table~\ref{tab:simulation-parameters}. We set the simulation time to
1000 seconds to ensure that the system runs long enough after reaching
a steady state. Based on a preliminary study, the system typically
reaches steady state after around 200 seconds. We ran 50 independent
replications for each parameter combination; and all shown results are
averaged over these 50 runs. The computed 95\% confidence intervals
are generally narrow, indicating our results are statistically stable
across these replications. These intervals are either shown as error
bars in the figures or omitted when the gap is small.

\begin{table}[h!]
  \centering
    \caption{Fixed Simulation Parameters}
  \renewcommand{\arraystretch}{1.5}
  \begin{tabular}{|l|l|}
    \hline
    \textbf{Parameter}                  & \textbf{Value}                     \\ \hline
    INET Version                        & 4.2.5                               \\ \hline
    OMNeT++ Version                     & 5.6.2                              \\ \hline
    Path Loss Exponent                  & 2                                  \\ \hline
    Thermal Noise Power                 & -110 dBm                            \\  \hline
    Neighbor Table Timeout                        & 10 s                              \\ \hline
    Minimal Safety Distance             & 0.5 m                              \\ \hline
    Radius                              & 10 m                               \\ \hline
    Speed Interval                      & 10--15 m/s                         \\ \hline
    Speed Distribution                  & Uniform                            \\ \hline
    Drone inter-generation time distribution       & Exponential                 \\ \hline
    Entry/Exit Lanes                    & 21                                 \\ \hline
    Replication Number per Simulation   & 50                                 \\ \hline
    Algorithm Period (Normal)           & 2 s                                \\ \hline
    Algorithm Period (Stop)             & 0.5 s                              \\ \hline
    Simulation time                     & 1000 s                             \\
    \hline
  \end{tabular}
  \label{tab:simulation-parameters}
\end{table}

The fixed algorithm parameters are listed in
Table~\ref{tab:algorithm-parameters}. We do not vary $\epsilon_0$, as
it should be a function of the maximum allowed speed and deceleration
(this is the distance it should take for a drone at maximum speed to
slow down to zero speed). The same applies to
$\epsilon_2$. Considering the maximum drone speed (15 m/s) and the
default algorithm period (2 seconds), both $\epsilon_2$ and
$\epsilon_0$ are set to 30 meters. This configuration ensures that
drones do not enter closed areas or experience direct rear-end
collisions with leading drones. We also do not vary $\epsilon_3$ as
this parameter only applies in particular situations, predominantly
when the current drone is stationary around its target point or when a
nearby neighbor drone is in very close proximity and moving at a
higher speed. In the first case, the minimum required value of
$\epsilon_3$ that can theoretically avoid all collisions is
0.75 meters, this is given by:
\begin{equation}
    \frac{\epsilon_3*\frac{\mbox{radius}}{s}}{\vmax} \geq \mbox{minimal safety distance},
\end{equation}
where the left-hand term represents the distance to the neighboring
drone in the worst case, which should be greater than the minimum
safety distance. In the second case $\epsilon_3$ should be larger than
2.5 meters, this value is derived as follows:
\begin{equation}
    \epsilon_3 \geq (\vmax-\vmin) * \frac{s}{2},
\end{equation}
where $\vmin = 10$ m/s, based on our selected speed interval in
Table~\ref{tab:simulation-parameters}. This inequality ensures that
the current drone does not initiate a switching during the first half
of a nearby neighboring drone’s switching process,\footnote{In the
second half of the switching process, the nearby drone is considered
to belong to its destination lane. As a result, the current drone is
also prohibited from starting a switch to this lane, according to our
collision cost function.} to avoid the potential collisions. Combining
these two situations, we finally set $\epsilon_3$ to 3 meters, which
includessome safety margin (e.g.\ to accommodate packet loss).

As a result, our remaining study focuses on the two MAC / PHY
parameters beacon rate and transmit power, and the three algorithm
parameters $\kappa_1$, $\kappa_2$ and $\epsilon_1$. All these are
investigated for varying generation rate.

\begin{table}[ht!]
  \centering
    \caption{Fixed Algorithm Parameters}
  \renewcommand{\arraystretch}{1.5}
  \begin{tabular}{|l|l|l|}
    \hline
    \textbf{Parameter}              & \textbf{Notation}    & \textbf{Value}                     \\ \hline
    Algorithm Period (Normal)       & $u_{normal}$    & 2 s                                \\ \hline
    Algorithm Period (Stop)         & $u_{stop}$    & 0.5 s                              \\ \hline
    Time for lane switching       & $s$    & 1 s                                \\ \hline
    Maximum Cost Definition         & $C_{max}$    & 30000                  \\ \hline
        Distance Threshold for the Closing Point & $\epsilon_0$    &  30 m  \\ \hline
    Distance Threshold for Searching Ahead Drone  & $\epsilon_2$    &   30 m  \\ \hline
    Distance Threshold for the Switching Priority & $\epsilon_3$    &    3 m  \\ \hline
  \end{tabular}
  \label{tab:algorithm-parameters}
\end{table}

\subsection{Key Performance Metrics}
\label{sec:results:key-metrics}

We use three key performance metrics, which are:
\begin{itemize}
\item \textbf{Collision Rate (cr)}: This is given as the ratio of
      collided drones to all injected drones. It directly reflects how
      safe the drones are while cruising under the STDG algorithm.
\item \textbf{Average Speed (as)}: Calculated as the total flight
      distance divided by the total active time of all drones. Without
      any interference, this value should be 12.5 m/s, based on the
      selected speed interval. A higher average speed generally
      indicates that drones experience less speed loss under the STDG
      algorithm.
\item \textbf{Throughput}: This is the rate at which drones reach
      their destinations. In our simulations, we specifically track
      the following three metrics over time:
      \begin{itemize}
      \item \textbf{Injection count (ic)}: the number of nodes
            actually injected into the system during each second.
      \item \textbf{Arrival count (ac)}: the number of nodes that
            reach their destinations during each second.
      \item \textbf{Total Node count (nc)}: the total number of active
            drones (drones currently in the system) during each
            second.
      \end{itemize} Among these, the arrival count represents the
      actual throughput observed in the system. A higher arrival count
      indicates that the system can accommodate more drones and
      complete more deliveries within the same time interval. However,
      the upper limit of the arrival count is constrained by the
      injection count. The total node count is another parameter that
      influences the arrival count: when the system include more
      active drones, traffic congestion will become more frequent,
      leading to a lower arrival count.
\end{itemize}

\subsection{Initial Study: Evaluation of Beacon Rate and Transmit Power}
\label{sec:results:initial-study}

In the initial study, we explore the impact of two key MAC/PHY
parameters (beacon rate, transmit power) and select suitable
combinations of these parameters for our further analysis of algorithm
parameters. The values we used for these two parameters are listed in
Table~\ref{tab:simulation-configuration}, the fixed values of the
algorithm parameters are given in
Table~\ref{tab:algorithm-parameters-inital-study}. These algorithm
parameters have been chosen to make drone behavior more conservative,
with drones only deciding to switch when no neighboring drones are
nearby.

\begin{table}[ht!]
  \centering
    \caption{Variable Parameters in Initial Study}
  \renewcommand{\arraystretch}{1.5}
  \begin{tabular}{|m{3cm}|m{3cm}|m{7cm}|}
    \hline
    \textbf{Parameter}            & \textbf{Values}  &\textbf{Description}                                       \\ \hline
    Generation Rate               & 0.05, 0.1, 0.2 & Drones injected per second per lane       \\ \hline
    Beaconing Interval            & 5 Hz, 20 Hz, 50 Hz & Beaconing frequency per drone        \\ \hline
    Transmit Power                & 2 mW, 20 mW & Signal power for
    beaconing packets, results in reception/interference ranges of
    250\,m / 4445\,m and 791\,m / 14058\,m, respectively      \\ \hline
  \end{tabular}
  \label{tab:simulation-configuration}
\end{table}

\begin{table}[ht!]
  \centering
    \caption{Other Algorithm Parameters in Initial Study}
  \renewcommand{\arraystretch}{1.5}
  \begin{tabular}{|l|l|l|}
    \hline
    \textbf{Parameter}              & \textbf{Notation}    & \textbf{Value}                     \\ \hline
    Position Cost Coefficient         & $\kappa_1$    & 10                  \\ \hline
    Collision Cost Coefficient         & $\kappa_2$    &  30000             \\ \hline
    Distance Threshold for Searching Neighbor Drone &   $\epsilon_1$    &  15 m  \\ \hline
  \end{tabular}
  \label{tab:algorithm-parameters-inital-study}
\end{table}

\begin{figure}[htbp]
  \centering
  \includegraphics[scale=0.4]{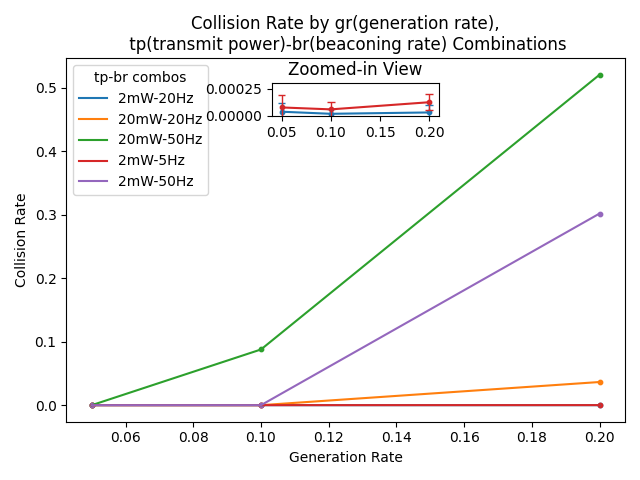}
  \caption{Collision Rate for Different Transmit Power and Beaconing Rate}
  \label{fig:collision-rate-initial}
\end{figure}

Figure~\ref{fig:collision-rate-initial} shows the average collision
rate for all considered combinations of beacon rate and transmit
power. The maximum collision rate observed over all replications is
0.567, obtained when choosing a 50\,Hz beaconing rate and 20\,mW
transmit power. While collisions happen in all scenarios, their rate
is particularly high when the transmit power is 20\,mW or the
beaconing frequency is 50\,Hz. For other parameter combinations, the
95\% confidence intervals, shown in the zoomed-in sub-figure, indicate
very low collision rates even under the highest drone densities.

\begin{figure}[htbp]
  \centering
  \includegraphics[scale=0.4]{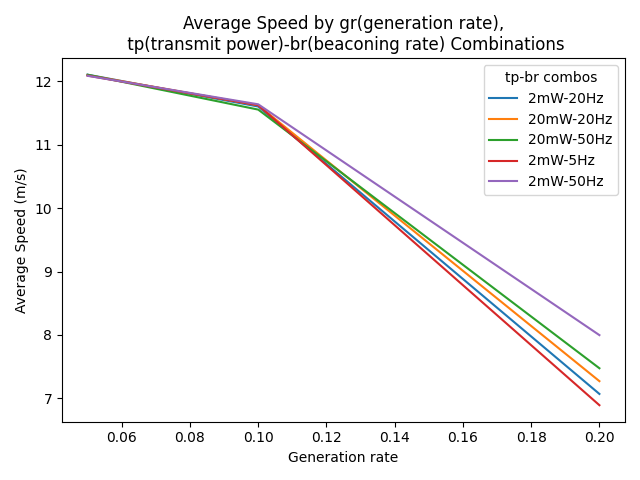}
  \caption{Average Speed for Different Parameter Combinations}
  \label{fig:average-speed}
\end{figure}

Figure~\ref{fig:average-speed} presents the average speed results for
the considered parameter combinations. It also shows the speed loss
(the difference between the expected average speed and the actual
average speed) of drones as the expected average speed is 12.5\,m/s
due to our configuration for $\vpref$. There are significant
differences between the lines, the 50\,Hz cases result in the highest
average speed. Comparing this finding with the collision rate data in
Figure~\ref{fig:collision-rate-initial} suggests that a higher
collision rate leads to a faster average speed, presumably since
collisions reduce the drone density in the system, which in turn
allows the remaining drones to move at higher speeds more often.

\begin{figure}[htbp]
  \centering
  \includegraphics[scale=0.4]{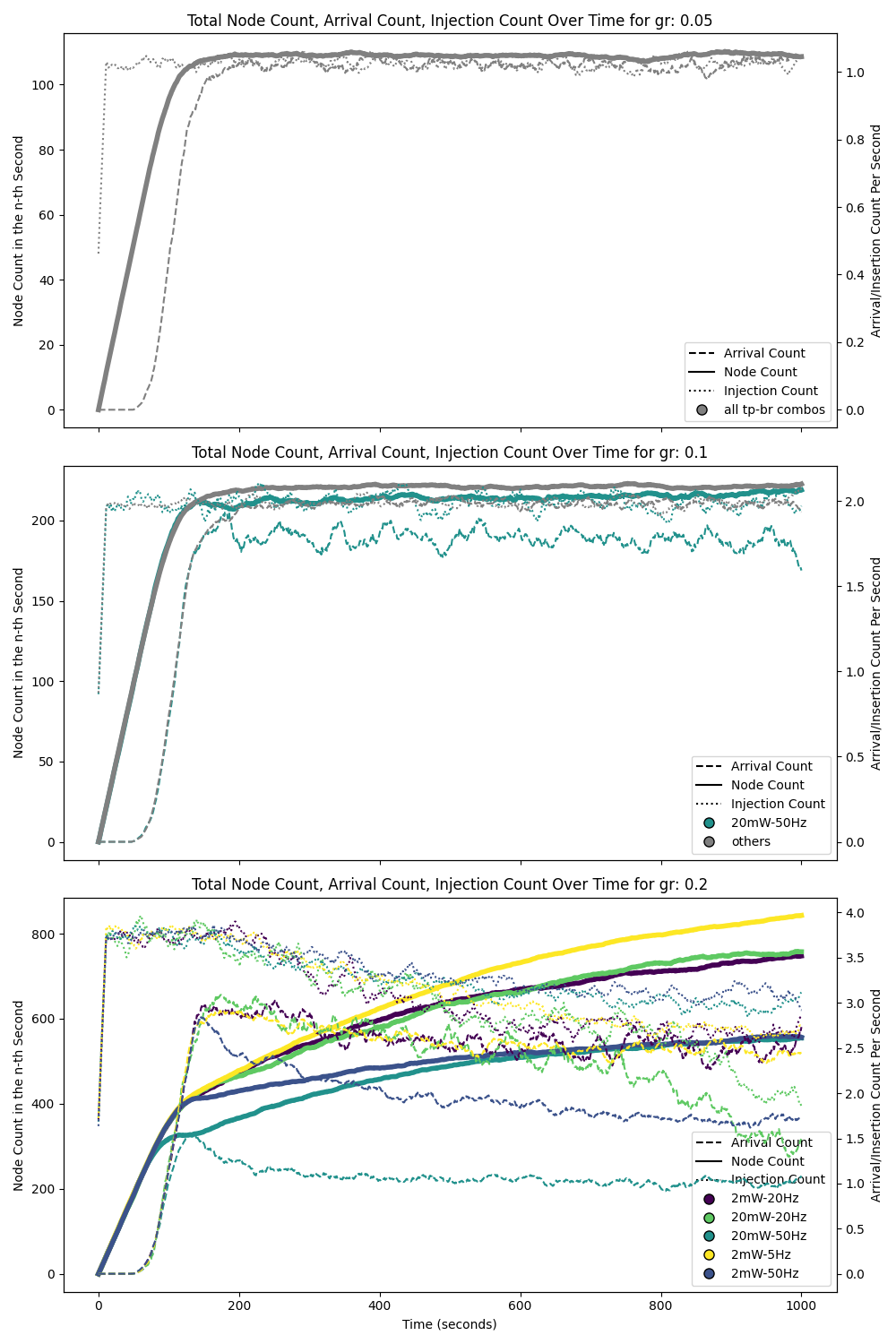}
  \caption{Throughput under different situation}
  \label{fig:node-count}
\end{figure}

Figure \ref{fig:node-count} illustrates the detailed trends of node
injection/arrival counts and the total node count over time for three
different values of the generation rate and selected parameter
combinations. In all subfigures the x-axis represents the simulation
time in seconds. The left y-axis corresponds to the node count shown
by the solid line, while the right y-axis represents the injection and
arrival counts, indicated by two types of dashed lines. Note that, in
most cases (except gr=0.2), running the simulation for 200 seconds is
sufficient for the system to reach the steady state. For gr=0.2, that
process becomes much slower, even 1000 seconds is not enough. But we
can still observe that the slopes of the curves begin to flatten over
time.

In the first two sub-figures ($gr = 0.05$ and $0.1$), all parameter
combinations result in similar performance. However, when $gr = 0.2$,
significant differences appear. The combinations 2\,mW / 20\,Hz and
2\,mW / 5\,Hz achieve the best throughput, exhibiting the highest
total node count and arrival count.

Parameter sets with a 50\,Hz beaconing rate result in lower total node
counts, while those with 20\,mW transmit power show lower arrival
counts. Based on these observations, we conclude that in high-density
drone environments, the 2\,mW / 20\,Hz and 2\,mW / 5\,Hz
configurations are more suitable for maintaining good throughput
performance under our algorithm.

Combining the above observation, the 2\,mW / 20\,Hz and 2\,mW / 5\,Hz
combinations result in the best performance in terms of collision
rate, average speed, and throughput. Since 10\,Hz lies between 20\,Hz
and 5\,Hz, and is also a commonly used configuration in vehicular
ad-hoc networks \cite{karagiannis2011vehicular}, we finally select the
2\,mW / 10\,Hz combination as the representative configuration for our
subsequent simulations.

\subsection{Sensitivity Study: Impact of Algorithm Parameters}
\label{sec:results:sensitivity-study}

Our STDG algorithm has a number of adjustable parameters ($\kappa_1$,
$\kappa_2$, and $\epsilon_2, \ldots, \epsilon_3$) and in this section
we explore the relative impact of some of these parameters on our
chosen performance metrics.

\subsubsection{Sensitivity Analysis Method}
\label{sec:results:sensitivity:method}

We quickly describe the method used for sensitivity analysis in this
paper, which is similar to the method used in
\cite{Moravejosharieh:Willig:15} and \cite{Willig:25} and based on the
response surface methodology \cite{Myers:Montgomery:02},
\cite{Jain:91}, which attempts to explain the response variables (the
performance metrics described in
Section~\ref{sec:results:key-metrics}) in terms of a regression model,
which in this paper has been chosen to be a second-order polynomial
model -- hence a response variable $Y$ depends on the $k$ factors
$x_i$ like:
\begin{equation}
  \label{eq:regression-ansatz}
  Y = \alpha_0 + \sum_{i=1}^k \alpha_i x_i
     + \sum_{i=1}^k \sum_{j<i} \alpha_{i,j} x_i  x_j
\end{equation}
For each combination of factor values $\vect{x} = (x_1,\ldots,x_k) \in
\set{-1,1}^k$ the response value $y_{\vect{x}}$ is obtained as an
average over 50 simulation replications. The model parameters
$\alpha_0$ (known as the intercept), $\alpha_i$ (linear coefficients)
and $\alpha_{i,j}$ (mixed coefficients or interactions) are then
obtained from a least-squares regression. Next we calculate the
sum-of-squares-total (SST)
\begin{equation}
  \label{eq:sst}
  SST = \sum_{\vect{x} \in \set{-1,1}^k} (y_x - \bar{y})^2
\end{equation}
and the sum-of-squares-errors
\begin{eqnarray}
  \label{eq:sse}
  \lefteqn{SSE = } \\
  \nonumber & & \sum_{\vect{x} \in \set{-1,1}^k} \left(y_x - \left(
          \alpha_0 + \sum_{i=1}^k \alpha_i x_i
     + \sum_{i=1}^k \sum_{j<i} \alpha_{i,j}  x_i  x_j\right)\right)^2
\end{eqnarray}
giving the total squared error between the regression model and
observed responses. The SST can be represented in terms of the model
parameters as
\begin{equation}
  SST = 2^k (\alpha_1^2 + \ldots + \alpha_k^2 + \alpha_{2,1}^2
  + \alpha_{3,1}^2
  + \alpha_{3,2}^2 +
  \ldots + \alpha_{k,k-1}^2),
\end{equation}
and the contribution or relative impact of factor $k$ or any
interaction between factors $k$ and $j$ to the overall SST is then:
\begin{equation}
  \label{eq:impact-of-factors}
  \frac{2^k \alpha_k^2}{SST}  \qquad , \qquad \frac{2^k \alpha_{k,j}^2}{SST}.
\end{equation}
To assess how well the regression model fits the data we use the
coefficient of determination (or $R^2$ value):
\begin{equation}
  \label{eq:coefficient-of-determination}
  R^2 = \frac{SST-SSE}{SST}.
\end{equation}
The $R^2$ value expresses the total variation explainable by the
regression model, values close to one indicate a very good
representation of the data by the model.


\begin{table}[ht!]
  \centering
    \caption{Algorithm Parameters Varied in Sensitivity Study}
  \renewcommand{\arraystretch}{1.5}
  \begin{tabular}{|l|l|l|}
    \hline
    \textbf{Parameter}    & \textbf{Symbol}    &   \textbf{Values}  \\
    \hline
    Position Cost Coefficient    & $\kappa_1$    & \set{1, 100}  \\
    \hline
    Collision Cost Coefficient    & $\kappa_2$    &  \set{1, 100}  \\
    \hline
    Distance Threshold for Searching Neighbor Drone &   $\epsilon_1$  &  \set{5\,\mbox{m}, 25\,\mbox{m}}  \\
    \hline
    Generation rate  & gr & \set{0.05, 0.1, 0.2} \\
    \hline
  \end{tabular}
  \label{tab:algorithm-parameters-sensitivity-study}
\end{table}


\subsubsection{Considered Parameters and Results}

In our study we have varied the cost coefficients $\kappa_1$ and
$\kappa_2$, the distance threshold $\epsilon_1$ and the generation
rate, see Table~\ref{tab:algorithm-parameters-sensitivity-study} for
the values considered. Table \ref{tab:fa-results} summarizes the
sensitivity study results, including the $R^2$ values, regression
coefficients, and percentage contributions of each parameter. The
models achieved a good fit to the data, with $R^2$ values of 92.39\%,
95.43\%, and 98.53\% for collision rate, average speed, and arrival
count, respectively.

\begin{table}
    \caption{Factor analysis results}
  \begin{center}
    \begin{tabular}{|l|l|l|l|}
   \hline 
   & \textbf{Collision rate}  & \textbf{Average speed (m/s)} & \textbf{Arrival count}\\ 
   \hline 
   Average of responses &  7.7e-3 &  10.9419 &  2.4369 \\ 
   \hline 
   Minimum of responses &  0.0 &  5.5121 &  1.32 \\ 
   \hline 
   Maximum of responses &  2.58e-2 &  12.4319 &  3.79 \\ 
   \hline 
   $R^2$ value (\%) &  92.3857 &  95.4342 &  98.5268 \\ 
   \hline 
   \multicolumn{4}{|l|}{\textbf{Regression coefficients}} \\ 
   \hline 
   Cost parameter $\kappa_1$ (Coefficient $\alpha_1$) &  1.8e-3 &  0.5261 &  0.2919  \\ 
   \hline 
   Cost parameter $\kappa_2$ (Coefficient $\alpha_2$) &  -2.8e-3 &  -0.5083 &  4.19e-2 \\ 
   \hline 
   Distance threshold parameter $\epsilon_1$ (Coefficient $\alpha_3$) &  3.1e-3 &  -1.0931 &  -0.3944  \\ 
   \hline 
   Generation rate (Coefficient $\alpha_4$) &  3.7e-3 &  -0.7248 &  0.5981  \\ 
   \hline 
   \multicolumn{4}{|l|}{\textbf{Percentage contributions}} \\ 
   \hline 
   Cost parameter $\kappa_1$ (\% contr.) &  5.9603 &  8.9514 &  11.4595 \\ 
   \hline 
   Cost parameter $\kappa_2$ (\% contr.) &  13.8383 &  8.3539 &  0.2359 \\ 
   \hline 
   Distance threshold parameter $\epsilon_1$ (\% contr.) &  16.5963 &  38.644 &  20.9214 \\ 
   \hline 
   Generation rate (\% contr.) &  24.5403 &  16.9908 &  48.1234 \\ 
   \hline 
\end{tabular}

    \label{tab:fa-results}
  \end{center}
\end{table}

\begin{table}
    \caption{Effect of interaction terms in factor analysis}
  \begin{center}
\begin{tabular}{|l|l|l|l|}
\hline
 \textbf{Regression coefficients}   & \textbf{Collision rate}  & \textbf{Average speed (m/s)} & \textbf{Arrival count}\\ 
\hline
 $\kappa_1\times \kappa_2$ &  1.5e-3 &  0.2439 &  -0.0206  \\ 
\hline
 $\kappa_1\times \epsilon_1$ &  1.8e-3 &  0.4392 &  0.2381  \\ 
\hline
 $\kappa_1\times gr$ &  1.1e-3 &  0.1379 &  0.1256  \\ 
\hline
 $\kappa_2\times \epsilon_1$ &  -1.8e-3 &  -0.3103 &  -0.0219  \\ 
\hline
 $\kappa_2\times gr$ &  -1.4e-3 &  -0.1377 &  0.0306  \\ 
\hline
 $\epsilon_1\times gr$ &  2.4e-3 &  -0.5558 &  -0.2406  \\ 
\hline
 \textbf{Percentage contributions}   & \textbf{Collision rate}  & \textbf{Average speed (m/s)} & \textbf{Arrival count}\\ 
\hline
 $\kappa_1\times \kappa_2$ (\% contr.) &  4.1727 &  1.923 &  0.0572  \\ 
\hline
 $\kappa_1\times \epsilon_1$ (\% contr.) &  5.4898 &  6.2371 &  7.6275  \\ 
\hline
 $\kappa_1\times gr$ (\% contr.) &  2.129 &  0.6151 &  2.1229  \\ 
\hline
 $\kappa_2\times \epsilon_1$ (\% contr.) &  5.8839 &  3.1143 &  0.0644  \\ 
\hline
 $\kappa_2\times gr$ (\% contr.) &  3.4206 &  0.6132 &  0.1262  \\ 
\hline
 $\epsilon_1\times gr$ (\% contr.) &  10.3546 &  9.9914 &  7.7885  \\ 
\hline
\end{tabular}

    \label{tab:fa-results-interaction}
  \end{center}
\end{table}

Across all outputs, the generation rate showed the largest
contribution to arrival count (48.12\%) and collision rate
(24.54\%). The distance threshold $\epsilon_1$ contributed
significantly to average speed (38.64\%, negative) and collision rate
(16.60\%, positive). The other regression coefficients indicate that
$\kappa_1$ generally has a positive effect on all key metrics, while
$\kappa_2$ tends to have a negative effect on average speed and
collision rate.

Our factor analysis also reveals that there are interaction effects
between the parameters, the relevant statistics are shown in Table
\ref{tab:fa-results-interaction}. Among the six interaction terms, the
$\epsilon_1 \times gr$ term shows the largest influence on the
collision rate (10.35\%), average speed (9.99\%), and arrival count
(7.79\%). Other interaction terms involving $\epsilon_1$, such as
$\kappa_1 \times \epsilon_1$ and $\kappa_2 \times \epsilon_1$, also
have obvious effects, especially on average speed. In contrast,
interaction terms which do not involve $\epsilon_1$, such as $\kappa_1
\times \kappa_2$ or $\kappa_1 \times gr$, contribute less than 5\% to
most response metrics.

\begin{figure}[htbp]
  \centering
  \includegraphics[scale=0.65]{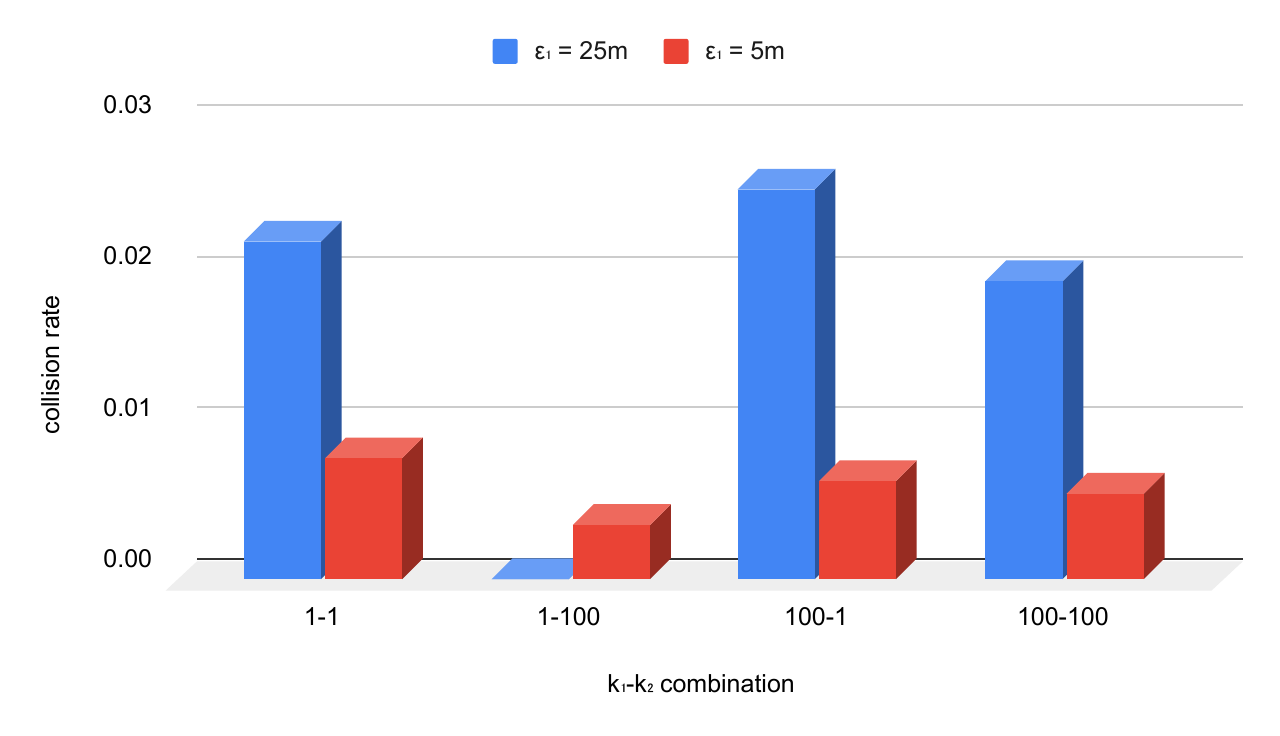}
  \caption{Collision Rate for Different $\kappa_1$, $\kappa_2$, and $\epsilon_1$ Combinations at gr = 0.2}
  \label{fig:cr-gr0.2}
\end{figure}

\begin{figure}[htbp]
  \centering
  \includegraphics[scale=0.65]{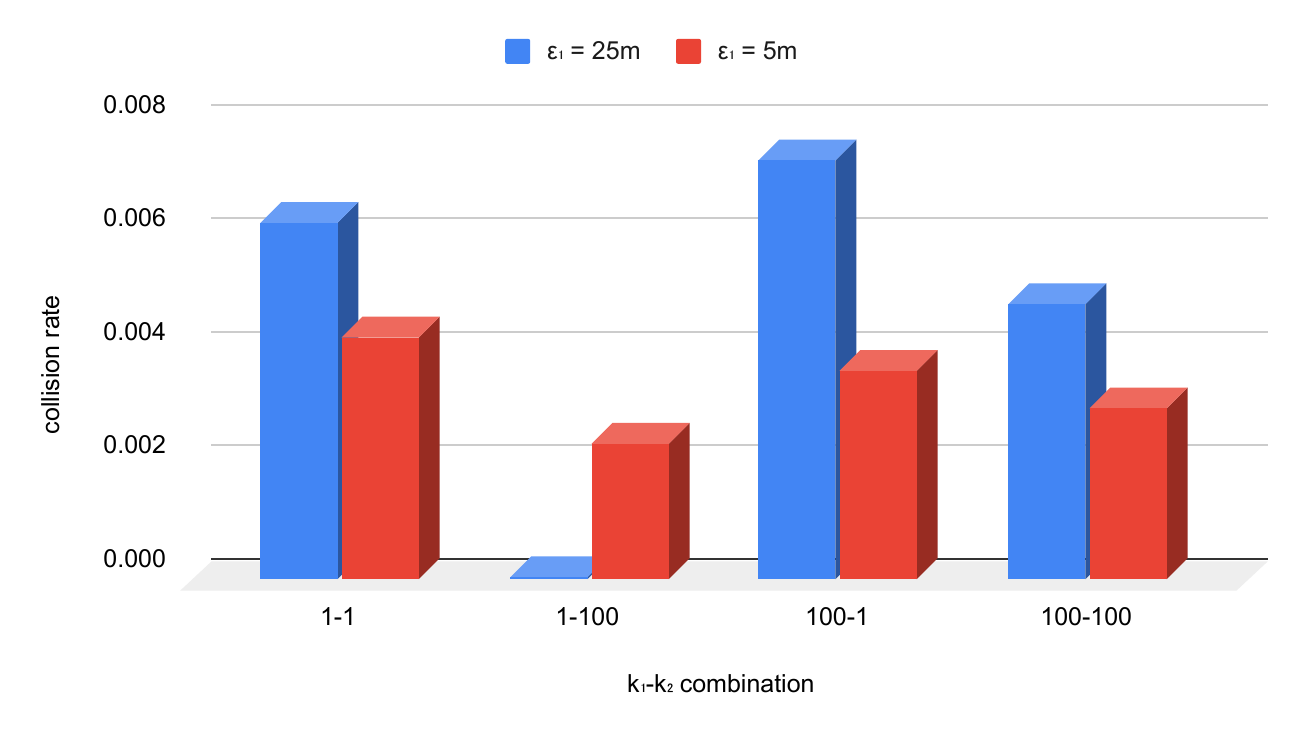}
  \caption{Collision Rate for Different $\kappa_1$, $\kappa_2$, and $\epsilon_1$ Combinations at gr = 0.1}
  \label{fig:cr-gr0.1}
\end{figure}

From the table, the collision rate appears to be influenced by all
interaction terms. To fully understand the combined effects and their
relative importance, additional results are provided for further
discussion. Figure \ref{fig:cr-gr0.2} and Figure \ref{fig:cr-gr0.1}
show the collision rates for different parameter combinations. When
$gr = 0.2$, increasing $\epsilon_1$ has a positive effect (i.e.,
increases the rate) on the collision rate, except for the 1 / 100
combination for $\kappa_1$ / $\kappa_2$. Similar observations can be
made when the generation rate is low. However, the positive effect
tends to be smaller in most cases, and for the 1 / 100 combination for
$\kappa_1$ / $\kappa_2$, the negative influence becomes more
pronounced.

\subsection{Discussion}
\label{sec:results:discussion}

This study was separated into two stages to systematically evaluate
the performance of the STDG algorithm. In the initial stage, we focus
on varying the wireless network parameters, such as transmission power
and beaconing rate, to identify the effects of these parameters and
find out the most suitable configuration. The results suggest that the
choice of wireless network parameters has a significant impact on all
key metrics, particularly under high drone density. A balanced
configuration, 10Hz beaconing rate with 2mW transmit power, was found
to provide an optimal performance in most cases. However, Figure
\ref{fig:average-speed} and Figure \ref{fig:collision-rate-initial}
indicate that there exists a trade-off between collision rate and
average speed, where a lower beaconing rate will lead to a relatively
higher average speed but at the cost of (and enabled by) a higher
collision rate. These two parameters together affect the packet loss
rate, which in turn influences position and velocity
uncertainties. This finding suggests a research direction in which
wireless communication parameters could be dynamically adapted to
lower the packet loss rate and thereby optimize algorithm performance
across different situations.

In the sensitivity analysis stage, we vary some key algorithm
parameters such as $\kappa_1$, $\kappa_2$, and $\epsilon_1$ to explore
the effects of these parameters to our key metrics. Based on the
results, several key findings can be discussed.

The range of responses across different parameter combinations is
substantial, highlighting the impact of algorithm parameters on
performance. Specifically, the collision rate varies from almost zero
to as high as 2.58\%, which means that under some parameter
combinations collisions can be almost fully avoided, while under
others they will become a frequent issue. The average speed ranges
from 5.51 m/s to 12.43 m/s, indicating that some combinations lead
drones to fly at near their preferred speed (12.5 m/s on average),
while many others force them to slow down to maintain safety. The
range of arrival count also shows significant variation, from 1.32 to
3.79, which means some combinations can help system maintain a high
level of throughput, but others may lead to inefficiency. These
findings suggest that selecting suitable parameter values is
important for balancing safety (collision rate), efficiency (average
speed), and throughput (arrival count).

The generation rate (gr) has the most significant impact on key
metrics, such as collision rate (24.54\% contribution) and arrival
count (48.12\% contribution). Higher generation rates increase drone
density, which increases the likelihood of collisions and can lead to
some level of traffic congestion, reducing the average speed. In
addition, a larger number of drones naturally increases the total
system population, resulting in higher arrival count and
throughput. Although generation rate is not an algorithm parameter, it
strongly affects the selection of algorithm parameters, especially
$\epsilon_1$. Among the interaction terms involving the generation
rate, the combination of $\epsilon_1$ and gr shows the largest
contribution to all key metrics, accounting for 10.35\% of collision
rate, 9.99\% of average speed, and 7.79\% of arrival count. This
indicates that the performance is sensitive to the interaction effect
between the distance threshold $\epsilon_1$ and drone generation
rate. For example, in high generation rate scenarios, a larger
$\epsilon_1$ allows drones to start lane switching more cautiously to
ensure safety, whereas in lower generation rate scenarios, a smaller
$\epsilon_1$ can help improve average speed and throughput.

The position cost coefficient $\kappa_1$ has positive effects on all
key metrics, making contributions of 5.96\%, 8.95\%, and 11.46\% to
collision rate, average speed, and arrival count,
respectively. Increasing the value of $\kappa_1$ will lead drones to
become more aggressive in switching lanes toward their targets,
resulting in a higher average speed and arrival count, but also a
higher collision rate. Similar to the generation rate, $\kappa_1$ also
has a noticeable interaction (5.49\% 6.24\%, and 7.63\%) with
$\epsilon_1$. This indicates that the impact of $\kappa_1$ on
performance depends on $\epsilon_1$, with their combination affecting
collision rate, average speed, and throughput.

Opposite to $\kappa_1$, the collision cost coefficient $\kappa_2$
generally has negative effects on all key metrics, contributing
13.84\%, 8.35\%, and 0.24\% to collision rate, average speed, and
arrival count, respectively. Increasing $\kappa_2$ encourages drones
to apply more conservative actions, avoiding lane switching when other
drones have similar intentions. This can effectively reduce collision
rate but also lowers average speed and throughput. The interaction
effects involving $\kappa_2$ are concentrated on collision rate
(13.48\% in total), while the impact on on average speed (5.65\%) and
arrival count (0.25\%) is relatively smaller, indicating that
$\kappa_2$ primarily influences safety. These results suggest that
careful selecting $\kappa_2$ is important to balance collision
avoidance and the delivery efficiency, and its effect is largely
independent of other parameters.

Among the algorithm parameters, the distance threshold $\epsilon_1$
has the largest effect on all metrics, especially the average speed
(38.64\% contribution). Recall that $\epsilon_1$ determines the
distance at which drones detect others as neighbors. Once other drones
are identified as neighbors, they are included in the collision cost
calculation, which then interacts with other types of costs to
influence the drone’s decision-making. As a result, $\epsilon_1$ shows
strong interaction effects with other algorithm parameters, such as
$\kappa_1$ and $\kappa_2$. These findings indicate that $\epsilon_1$
is the most sensitive parameter in our algorithm, and its selection
should be made carefully while considering the values of other
parameters.

From Table \ref{tab:fa-results}, increasing $\epsilon_1$ significantly
reduces the average speed and arrival count. This can be explained by
the fact that a larger $\epsilon_1$ blocks more lane-switching
decisions, causing drones to wait longer to find an
opportunity. However, the results also show that a larger $\epsilon_1$
has a positive effect on collision rate, which is contrary to our
expectations. We present the detailed collision rate data in Figures
\ref{fig:cr-gr0.2} and \ref{fig:cr-gr0.1}. They clearly show the
interaction between $\epsilon_1$ and other parameters. When
$\kappa_1/\kappa_2$ is very small, increasing $\epsilon_1$ has a
significant effect in reducing the collision rate, which can even go
down to close-to-zero levels. In other cases, however, increasing
$\epsilon_1$ generally leads to worse results for all metrics.

During the simulations, an inner lane vanishing issue was observed. It
occurs when the drone generation rate is set to a high value,
resulting in a large number of drones slowly building up at the
hotspots before a connecting ramp. In this situation, the inner lane
of the road is often left unoccupied, leading to inefficient lane
usage. For our tested drone road system, 4 inner lanes (with lane identifier (0,0)) out of a total of 21 lanes all appear this phenomenon. An inner lane connects to a higher number of neighboring lanes,
making it more affected by potential collisions and congestion
(which tends to increase their collision cost). When drones assess the
feasibility of neighbor lanes for switching, the inner lane's high
potential collision cost results in it being given the lowest priority
for lane selection. This is the primary cause of the inner-lane
vanishing problem. Setting a larger $\kappa_1/\kappa_2$ can avoid this
issue, but as discussed above, a smaller $\kappa_1/\kappa_2$ is
necessary to achieve a close-to-zero collision rate. Therefore, it is
worthy to solve this problem as it could further increase average
speed and throughput without sacrificing the collision rate. In our
coordination-free algorithm, we found no effective way to fully
utilize the inner lane without introducing additional collision
risks. As a result, addressing this issue will be a key problem when
designing the coordination-based algorithm in the future research.


\section{Conclusion and Future Works}
\label{sec:conclusion}

In this paper, we proposed a drone system description standard, which
consists of two main components: roads and ramps, both of which are
uni-directional. Similar to ground vehicle infrastructure, drone roads
consist of multiple lanes arranged in a hexagonal network. Ramps are
categorized into on/off-ramps, which allow drones to enter or exit a
road halfway, and connecting ramps, which link two different
roads. 

We then introduced the Short-Term Decentralized Greedy (STDG) guidance
algorithm for navigation and collision avoidance within this drone
road system. STDG is coordination-free and relies only on GPS and a
wireless communication module, making it compatible with most
drones. The algorithm uses a cost model to evaluate the risk of
various actions such as staying in the current lane or switching to
other lanes; drones always select the action with the lowest cost.

A simulation study was conducted to evaluate the performance of the
proposed guidance algorithm within the drone road system, measuring
key metrics including collision rate, average speed, and throughput
under various parameter combinations. In the first stage, we varied
only the communication parameters to identify the most suitable
combination of transmit power and beaconing rate. Based on the
results, a 10\,Hz beaconing rate with 2\,mW transmit power led to the
most balanced performance in our scenario. In the sensitivity study
stage, we varied several algorithm parameters to explore their
influence on our chosen metrics. The results clearly show both direct
and interaction effects of these parameters, among which the distance
threshold $\epsilon_1$ is the most sensitive. Its value should be
carefully selected in conjunction with the other parameters. In
addition, a smaller $\kappa_1/\kappa_2$ ratio is an important
condition for achieving a close-to-zero collision rate, although this
comes with a reduction in average speed and throughput.

One potential issue was observed during the simulation, which is
called inner lane vanishing. While we do not demonstrate this in full
generality in this paper, we do suspect that this problem cannot be
avoided within the confines of a coordination-free approach, in which
drones share only limited types of information with each other,
significantly restricting their decision-making. Based on these
findings, one possible direction for future work is to develop a
coordination-based version of the algorithm, which could improve road
utilization and reduce congestion by enabling more informed
decision-making among drones.

In addition, packet loss was identified as a key factor affecting
system performance in our simulation. If drones could adapt certain
parameters based on real-time network conditions to improve
communication quality, the overall system metrics could be further
improved. This suggests that the guidance algorithm could be extended
to dynamically adjust parameters such as transmit power, and beacon
rate in response to changing communication conditions. Besides the
wireless settings, algorithm parameters such as cost coefficients ($\kappa_1$ and $\kappa_2$) and distance threshold ($\epsilon_0$, $\epsilon_1$, ...) could also be adjusted dynamically based on the current traffic conditions to further enhance the throughput and average speed.


\section*{Acknowledgment}

This project is supported by University of Canterbury Aho Hinatore
Accelerator Scholarship.



\begin{appendices}
\section{Constructing a Parallel Curve}
\label{sec:app:constructing-parallel-curve}

In this appendix we discuss one method for constructing a parallel
curve.

We are given a curve $\gamma : [a,b] \subset \reals \mapsto
\reals^3$. For the real space $\reals^3$ we assume that we are given a
Cartesian $x/y/z$ coordinate system with known origin and orientation,
and the coordinates of the curve points are expressed with reference
to this coordinate system. We assume that the $z$-axis is orthogonal
to the surface of the earth.

On the highest level, for a given curve point
$\vect{p}=\gamma(s)$ our method first constructs an auxiliary basis
within the plane going through \vect{p} that is orthogonal to the
tangent of the curve $\vect{\gamma'(s)}$ (this plane, which we will
call the \textbf{normal plane}, is spanned by two orthogonal vectors
originating in \vect{p} and pointing to points in the plane), then we
calculate the orthogonal projection of the fixed vector $\vect{p}
- (0,0,1)^T$ onto the normal plane, and take this orthogonal
  projection
as one basis vector of the final frame. The second basis vector is
obtained as the cross-product between the tangent vector of the curve
at \vect{p} and this orthogonal projection. This method will not work
when the normal plane is parallel to the $x/y$-plane, as the
orthogonal projection will reduce to the null vector. We now discuss
the details:
\begin{itemize}
\item We are given a curve
      $\gamma(\cdot):[0,L(\gamma)]\mapsto\reals^3$, already in
      arc-length presentation, and a parameter $s\in[0, L(\gamma)]$
      with corresponding point $\vect{p}=\gamma(s)$ on the curve and
      tangent vector $\vect{v}=\gamma'(s)$. As we assume our curve to
      be regular, we have $\vect{v}\ne\vect{0}$ and for the
      remainder we will in fact assume that it is normalized,
      i.e.\ $\norm{\vect{v}}=\norm{\gamma'(s)}=1$.
  \item As a very first step, we simplify the problem by translating
        \vect{p} to the origin (this translation then is reversed at
        the end). We henceforth assume $\vect{p}=\vect{0}$.
  \item To express the normal plane through point $\vect{p}$ and
        orthogonal to the tangent vector \vect{v} we start by finding
        two auxiliary and orthogonal vectors \vect{w_1} and \vect{w_2}
        in the normal plane, which we do as follows:
        \begin{itemize}
        \item In general, a plane in three dimensions going through a
          point $\vect{p}=(x_1,y_1,z_1)^T$ and having normal vector
          $(A,B,C)$ (which in our case will be the tangent vector
          \vect{v}) can be described as the point set
          \begin{displaymath}
            \setof{(x,y,z)^T\in\reals^3}{A(x-x_1) + B(y-y_1) + C(z-z_1) = 0}
          \end{displaymath} As we assume our curve to be regular, at
          least one of the three numbers $A$, $B$ and $C$ is
          nonzero. We are hence seeking solutions of the linear system
          \begin{displaymath}
            0
            = (A,B,C) \cdot
            \left[
              \begin{pmatrix} x \\ y \\  z \end{pmatrix}
              - \begin{pmatrix} x_1 \\ y_1 \\  z_1 \end{pmatrix}
              \right]
            = (A,B,C) \cdot
            \left[
              \begin{pmatrix} x \\ y \\  z \end{pmatrix}
              - \vect{p}
              \right]
          \end{displaymath}
          or, after introducing the abbreviation $\omega = (A,B,C)\cdot\vect{p}$,
          \begin{displaymath}
            \omega = (A,B,C) \cdot \begin{pmatrix} x \\ y \\  z \end{pmatrix}
          \end{displaymath} For the sake of example, suppose $A\ne
          0$. Then, to find the first point in the normal plane, set
          $y=1$ and $z=0$ and solve for $x$ as
          \begin{displaymath}
            x = \frac{\omega - B}{A}
          \end{displaymath}
          Hence the first point (or basis vector in the normal plane)
          \vect{w_1} is given by
          \begin{displaymath}
            \vect{w_1} =
            \begin{pmatrix}
              \frac{\omega-B}{A} \\ 1 \\ 0
            \end{pmatrix}
          \end{displaymath} and the second point / basis vector can be
          constructed as the cross product $\vect{w_2} = \vect{v}
          \times \vect{w_1}$ between \vect{v} and \vect{w_1} (perhaps
          after normalizing the latter two), which guarantees that
          \vect{w_2} is orthogonal to \vect{w_1} and \vect{v}. The two
          vectors \vect{w_1} and \vect{w_2} are hence our auxiliary
          basis for the normal plane.
        \end{itemize}
      \item Recalling that temporarily $\vect{p}=\vect{0}$ and that
            the normal plane hence passes through the origin, we now
            calculate the orthogonal projection of the vector
            $\vect{z_0} = (0,0,1)^T$ onto the normal plane. This
            orthogonal projection is given by the vector
    \begin{displaymath}
      \vect{u_1} = \inner{\vect{z_0}}{\vect{w_1}}\cdot\vect{w_1}
      + \inner{\vect{z_0}}{\vect{w_2}}\cdot\vect{w_2}
    \end{displaymath}
  where \inner{\cdot}{\cdot} denotes the inner product of two
  vectors. We take (after normalizing) \vect{u_1} as the first basis
  vector of our target frame in the normal plane, and construct the
  second basis vector \vect{u_2} as the cross-product
  $\vect{u_2}=\vect{v}\times\vect{u_1}$.
  \item After having constructed the frame (\vect{u_1},\vect{u_2}) for
        our normal plane, we determine the point on the parallel curve
        to be given by (in this plane) $f_1 \cdot \vect{u_1} + f_2
        \cdot \vect{u_2}$ where the coefficients $f_1$ and $f_2$
        determine the relative position of the parallel point in the
        normal plane.
 \end{itemize}



\section{Explanation of Hop Distance Calculation in Hex Grid Coordinate System}
\label{sec:app:explanation-hop-distance}

In the Section \ref{sec:algorithm:notations}, we introduce a formula
  \begin{displaymath}
    \hopdistance{(i,j)}{(i',j)} = \left\{\begin{array}{r@{\quad:\quad}l}
      |i-i'| + |j-j'| & (i-i')\cdot(j-j') \ge 0 \\
      \max\set{|i-i'|, |j-j'|} & \mbox{otherwise}
    \end{array}
    \right.
  \end{displaymath}
, to calculate the hop distance from one point $(i, j)$ to another $(i^\prime, j^\prime)$. Here we propose the detailed explanation of this formula.

We begin by assuming the following fundamental properties of the hexagonal grid coordinate system are known:
\begin{itemize}
    \item Each point in the system has six neighbors at a hop distance of 1. For the origin $(0, 0)$, its neighbors are $(1, 0)$, $(0, 1)$, $(1, -1)$, $(-1, 1)$, $(-1, 0)$, and $(0, -1)$.
    \item The hop distance between two points $\hopdistance{(i,j)}{(i',j)}$ equals to $\hopdistance{(i-i^\prime,j-j^\prime)}{(0,0)}$. If one point is the origin and another point is $(i,j)$, we then simply the notation of hop distance to $|(i,j)|$ in the following content.
    \item The addition of two points follows a basic component-wise rule, which is $(i, j) + (i^\prime, j^\prime) = (i+i^\prime, j+j^\prime)$. The same rule applies to subtraction.
    \item Multiplying both coordinates of a point by a constant scales its hop distance (to the origin) by the same absolute factor, which is $|(a*i,a*j)|=|a|*|(i,j)|$.
    \item The equality of the hop distance addition $|(i, j)| + |(i', j')| = |(i + i', j + j')|$ applies if and only if the angle between the two vector $(i,j)$ and $(i', j')$ is less than $90^o$ (more strictly, equals to $0^o$ or $60^o$ in our scenario).
\end{itemize}

Then, to calculate the $|(i-i^\prime,j-j^\prime)|$ (which equals to $\hopdistance{(i,j)}{(i',j)}$), we can transform $|(i-i^\prime,j-j^\prime)|$ to the following forms, if $(i-i')\cdot(j-j') \ge 0$ which indicates the angle between vectors $(i-i',0)$ and $(0,j-j')$ is acute
  \begin{displaymath}
    |(i-i^\prime,j-j^\prime)| = |(i-i',0)| + |(0,j-j')|
  \end{displaymath}
The transformed form can be further simplified to $|i-i^\prime|*|(1,0)| + |j-j^\prime|*|(0,1)|$, respectively. From the basic knowledge, both $|(1,0)|$ and $|(0,1)|$ are 1, hence the final hop distance is $|i-i^\prime)|*1 + |j-j^\prime| * 1 = |i-i^\prime| + |j-j^\prime|$

If $(i-i')\cdot(j-j') < 0$, letting $a = \min\set{|i-i'|, |j-j'|}$, we can transform the distance to:
  \begin{displaymath}
    |(i-i^\prime,j-j^\prime)| = \left\{\begin{array}{r@{\quad:\quad}l}
      |(i-i^\prime,-i+i^\prime)| + |(0,j-j^\prime+i-i^\prime)| & a = |i-i^\prime| \\
      |(-j+j^\prime,j-j^\prime)| + |(i-i^\prime+j-j^\prime, 0)| & a = |j-j^\prime|
    \end{array}
    \right.
  \end{displaymath}
Taking the first case ($a = |i-i^\prime|$) as an example, the angle between the vector $(i-i^\prime,-i+i^\prime)$ and $(0,j-j^\prime+i-i^\prime)$ must be acute since $(i-i^\prime)*0+(-i+i^\prime)*(j-j^\prime+i-i^\prime) = -1*((i-i')\cdot(j-j') + (i-i^\prime)^2)$  = $-1 * [(i-i^\prime)((j-j^\prime)+(i-i^\prime))]$ is always positive\footnote{$(j-j^\prime)+(i-i^\prime)$ has the same sign with $(j-j^\prime)$ since $|i-i^\prime|$ < $|j-j^\prime|$}. The first term $|(i-i^\prime,-i+i^\prime)|$ can be written as $|i-i^\prime|*|(1,-1)| = |i-i^\prime|$. And for second term it is $|j-j^\prime+i-i^\prime| = |j-j^\prime| - |i-i^\prime|$ since $(i-i')\cdot(j-j') < 0$. Finally, the hop distance $|(i-i^\prime,j-j^\prime)|$ equals $|j-j^\prime|$ when $\min\set{|i-i'|, |j-j'|} = |i-i^\prime|$, or equals $|i-i^\prime|$ when $\min\set{|i-i'|, |j-j'|} = |j-j^\prime|$. In other words, it always equals to $\max\set{|i-i'|, |j-j'|}$ when $(i-i')\cdot(j-j') < 0$.

\end{appendices}
\bibliographystyle{ieeetr}
\bibliography{bio.bib} 

\begin{thebibliography}{10}

\bibitem{moody2021value}
J.~Moody, E.~Farr, M.~Papagelis, and D.~R. Keith, ``The value of car ownership and use in the united states,'' {\em Nature Sustainability}, vol.~4, no.~9, pp.~769--774, 2021.

\bibitem{johnson2017cars}
E.~Johnson, ``Cars and ground-level ozone: how do fuels compare?,'' {\em European Transport Research Review}, vol.~9, no.~4, pp.~1--13, 2017.

\bibitem{piotrowska2019assessment}
K.~Piotrowska, W.~Kruszelnicka, P.~Ba{\l}dowska-Witos, R.~Kasner, J.~Rudnicki, A.~Tomporowski, J.~Flizikowski, and M.~Opielak, ``Assessment of the environmental impact of a car tire throughout its lifecycle using the lca method,'' {\em Materials}, vol.~12, no.~24, p.~4177, 2019.

\bibitem{dukkanci2024facility}
O.~Dukkanci, J.~F. Campbell, and B.~Y. Kara, ``Facility location decisions for drone delivery with riding: A literature review,'' {\em Computers \& Operations Research}, p.~106672, 2024.

\bibitem{li2023drone}
X.~Li, J.~Tupayachi, A.~Sharmin, and M.~Martinez~Ferguson, ``Drone-aided delivery methods, challenge, and the future: A methodological review,'' {\em Drones}, vol.~7, no.~3, p.~191, 2023.

\bibitem{eskandaripour2023last}
H.~Eskandaripour and E.~Boldsaikhan, ``Last-mile drone delivery: Past, present, and future,'' {\em Drones}, vol.~7, no.~2, p.~77, 2023.

\bibitem{rajabi2022drone}
M.~S. Rajabi, P.~Beigi, and S.~Aghakhani, ``Drone delivery systems and energy management: A review and future trends,'' {\em arXiv preprint arXiv:2206.10765}, 2022.

\bibitem{qu2022sensorless}
Z.~Qu and A.~Willig, ``Sensorless and coordination-free lane switching on a drone road segment—a simulation study,'' {\em Drones}, vol.~6, no.~12, p.~411, 2022.

\bibitem{cheng2015aerial}
E.~Cheng, {\em Aerial photography and videography using drones}.
\newblock San Francisco, USA: Peachpit Press, 2015.

\bibitem{puttock2015aerial}
A.~Puttock, A.~Cunliffe, K.~Anderson, and R.~E. Brazier, ``Aerial photography collected with a multirotor drone reveals impact of eurasian beaver reintroduction on ecosystem structure,'' {\em Journal of Unmanned Vehicle Systems}, vol.~3, no.~3, pp.~123--130, 2015.

\bibitem{varghese2017power}
A.~Varghese, J.~Gubbi, H.~Sharma, and P.~Balamuralidhar, ``Power infrastructure monitoring and damage detection using drone captured images,'' in {\em 2017 International Joint Conference on Neural Networks (IJCNN)}, (Anchorage, AK, USA), pp.~1681--1687, IEEE, 2017.

\bibitem{flammini2016railway}
F.~Flammini, C.~Pragliola, and G.~Smarra, ``Railway infrastructure monitoring by drones,'' in {\em 2016 International Conference on Electrical Systems for Aircraft, Railway, Ship Propulsion and Road Vehicles \& International Transportation Electrification Conference (ESARS-ITEC)}, (Toulouse, France), pp.~1--6, IEEE, 2016.

\bibitem{rosser2018surgical}
J.~C. Rosser~Jr, V.~Vignesh, B.~A. Terwilliger, and B.~C. Parker, ``Surgical and medical applications of drones: A comprehensive review,'' {\em JSLS: Journal of the Society of Laparoendoscopic Surgeons}, vol.~22, no.~3, 2018.

\bibitem{claesson2017time}
A.~Claesson, A.~B{\"a}ckman, M.~Ringh, L.~Svensson, P.~Nordberg, T.~Dj{\"a}rv, and J.~Hollenberg, ``Time to delivery of an automated external defibrillator using a drone for simulated out-of-hospital cardiac arrests vs emergency medical services,'' {\em Jama}, vol.~317, no.~22, pp.~2332--2334, 2017.

\bibitem{kumar2021novel}
A.~Kumar, R.~Krishnamurthi, A.~Nayyar, A.~K. Luhach, M.~S. Khan, and A.~Singh, ``A novel software-defined drone network (sddn)-based collision avoidance strategies for on-road traffic monitoring and management,'' {\em Vehicular Communications}, vol.~28, p.~100313, 2021.

\bibitem{outay2020applications}
F.~Outay, H.~A. Mengash, and M.~Adnan, ``Applications of unmanned aerial vehicle (uav) in road safety, traffic and highway infrastructure management: Recent advances and challenges,'' {\em Transportation research part A: policy and practice}, vol.~141, pp.~116--129, 2020.

\bibitem{nasa2015air}
U.~NASA, ``Air traffic management for low-altitude drones, na a,'' {\em SA (NASA), Washington DC, USA}, 2015.

\bibitem{abeyratne2014convention}
R.~Abeyratne, ``Convention on international civil aviation,'' {\em A Commentary, Switzerland}, 2014.

\bibitem{jang2017concepts}
D.-S. Jang, C.~A. Ippolito, S.~Sankararaman, and V.~Stepanyan, ``Concepts of airspace structures and system analysis for uas traffic flows for urban areas,'' in {\em AIAA Information Systems-AIAA Infotech@ Aerospace}, p.~0449, American Institute for Aeronautics and Astronautics (AIAA), 2017.

\bibitem{ijgi10050338}
D.~D. Nguyen, J.~Rohacs, and D.~Rohacs, ``Autonomous flight trajectory control system for drones in smart city traffic management,'' {\em ISPRS International Journal of Geo-Information}, vol.~10, no.~5, 2021.

\bibitem{aerospace8020038}
M.~Doole, J.~Ellerbroek, V.~L. Knoop, and J.~M. Hoekstra, ``Constrained urban airspace design for large-scale drone-based delivery traffic,'' {\em Aerospace}, vol.~8, no.~2, 2021.

\bibitem{svaigen2022biomixd}
A.~R. Svaigen, A.~Boukerche, L.~B. Ruiz, and A.~A. Loureiro, ``Biomixd: A bio-inspired and traffic-aware mix zone placement strategy for location privacy on the internet of drones,'' {\em Computer Communications}, vol.~195, pp.~111--123, 2022.

\bibitem{9256627}
B.~Pang, W.~Dai, T.~Ra, and K.~H. Low, ``A concept of airspace configuration and operational rules for uas in current airspace,'' in {\em 2020 AIAA/IEEE 39th Digital Avionics Systems Conference (DASC)}, pp.~1--9, 2020.

\bibitem{9945667}
Q.~Shao, R.~Li, M.~Dong, and C.~Song, ``An adaptive airspace model for quadcopters in urban air mobility,'' {\em IEEE Transactions on Intelligent Transportation Systems}, pp.~1--10, 2022.

\bibitem{hartenstein2009vanet}
H.~Hartenstein and K.~Laberteaux, {\em VANET: vehicular applications and inter-networking technologies}.
\newblock John Wiley \& Sons, 2009.

\bibitem{wischhof2005congestion}
L.~Wischhof and H.~Rohling, ``Congestion control in vehicular ad hoc networks,'' in {\em IEEE International Conference on Vehicular Electronics and Safety, 2005.}, pp.~58--63, IEEE, 2005.

\bibitem{fallah2010analysis}
Y.~P. Fallah, C.-L. Huang, R.~Sengupta, and H.~Krishnan, ``Analysis of information dissemination in vehicular ad-hoc networks with application to cooperative vehicle safety systems,'' {\em IEEE Transactions on Vehicular Technology}, vol.~60, no.~1, pp.~233--247, 2010.

\bibitem{zhang2008congestion}
W.~Zhang, A.~Festag, R.~Baldessari, and L.~Le, ``Congestion control for safety messages in vanets: Concepts and framework,'' in {\em 2008 8th International Conference on ITS Telecommunications}, pp.~199--203, IEEE, 2008.

\bibitem{wei2019identifying}
L.~J. Wei and J.~M.-Y. Lim, ``Identifying transmission opportunity through transmission power and bit rate for improved vanet efficiency,'' {\em Mobile Networks and Applications}, vol.~24, pp.~1630--1638, 2019.

\bibitem{1311801}
J.~Yin, X.~Wang, and D.~Agrawal, ``Optimal packet size in error-prone channel for ieee 802.11 distributed coordination function,'' in {\em 2004 IEEE Wireless Communications and Networking Conference (IEEE Cat. No.04TH8733)}, vol.~3, pp.~1654--1659 Vol.3, 2004.

\bibitem{chaabouni2013collision}
N.~Chaabouni, A.~Hafid, and P.~K. Sahu, ``A collision-based beacon rate adaptation scheme (cba) for vanets,'' in {\em 2013 IEEE International Conference on Advanced Networks and Telecommunications Systems (ANTS)}, pp.~1--6, IEEE, 2013.

\bibitem{gupta2015survey}
L.~Gupta, R.~Jain, and G.~Vaszkun, ``Survey of important issues in uav communication networks,'' {\em IEEE communications surveys \& tutorials}, vol.~18, no.~2, pp.~1123--1152, 2015.

\bibitem{hota2022performance}
L.~Hota, B.~P. Nayak, A.~Kumar, B.~Sahoo, and G.~M.~N. Ali, ``A performance analysis of vanets propagation models and routing protocols,'' {\em Sustainability}, vol.~14, no.~3, p.~1379, 2022.

\bibitem{9256561}
S.~James, R.~Raheb, and A.~Hudak, ``Impact of packet loss to the motion of autonomous uav swarms,'' in {\em 2020 AIAA/IEEE 39th Digital Avionics Systems Conference (DASC)}, pp.~1--9, 2020.

\bibitem{9624939}
L.~Ruan, G.~Li, W.~Dai, S.~Tian, G.~Fan, J.~Wang, and X.~Dai, ``Cooperative relative localization for uav swarm in gnss-denied environment: A coalition formation game approach,'' {\em IEEE Internet of Things Journal}, vol.~9, no.~13, pp.~11560--11577, 2022.

\bibitem{aksjonov2021rule}
A.~Aksjonov and V.~Kyrki, ``Rule-based decision-making system for autonomous vehicles at intersections with mixed traffic environment,'' in {\em 2021 IEEE International Intelligent Transportation Systems Conference (ITSC)}, pp.~660--666, IEEE, 2021.

\bibitem{7995949}
C.~Hubmann, M.~Becker, D.~Althoff, D.~Lenz, and C.~Stiller, ``Decision making for autonomous driving considering interaction and uncertain prediction of surrounding vehicles,'' in {\em 2017 IEEE Intelligent Vehicles Symposium (IV)}, pp.~1671--1678, 2017.

\bibitem{9147837}
S.~Bae, D.~Saxena, A.~Nakhaei, C.~Choi, K.~Fujimura, and S.~Moura, ``Cooperation-aware lane change maneuver in dense traffic based on model predictive control with recurrent neural network,'' in {\em 2020 American Control Conference (ACC)}, pp.~1209--1216, 2020.

\bibitem{9726894}
S.~Li, C.~Wei, and Y.~Wang, ``Combining decision making and trajectory planning for lane changing using deep reinforcement learning,'' {\em IEEE Transactions on Intelligent Transportation Systems}, vol.~23, no.~9, pp.~16110--16136, 2022.

\bibitem{wang2019lane}
J.~Wang, Q.~Zhang, D.~Zhao, and Y.~Chen, ``Lane change decision-making through deep reinforcement learning with rule-based constraints,'' in {\em 2019 International Joint Conference on Neural Networks (IJCNN)}, pp.~1--6, IEEE, 2019.

\bibitem{xu2022integrated}
C.~Xu, W.~Zhao, J.~Liu, C.~Wang, and C.~Lv, ``An integrated decision-making framework for highway autonomous driving using combined learning and rule-based algorithm,'' {\em IEEE Transactions on Vehicular Technology}, vol.~71, no.~4, pp.~3621--3632, 2022.

\bibitem{suh2018stochastic}
J.~Suh, H.~Chae, and K.~Yi, ``Stochastic model-predictive control for lane change decision of automated driving vehicles,'' {\em IEEE Transactions on Vehicular Technology}, vol.~67, no.~6, pp.~4771--4782, 2018.

\bibitem{farid2022modified}
G.~Farid, S.~Cocuzza, T.~Younas, A.~A. Razzaqi, W.~A. Wattoo, F.~Cannella, and H.~Mo, ``Modified a-star (a*) approach to plan the motion of a quadrotor uav in three-dimensional obstacle-cluttered environment,'' {\em Applied Sciences}, vol.~12, no.~12, p.~5791, 2022.

\bibitem{wang2022trajectory}
J.~Wang, Y.~Li, R.~Li, H.~Chen, and K.~Chu, ``Trajectory planning for uav navigation in dynamic environments with matrix alignment dijkstra,'' {\em Soft Computing}, vol.~26, no.~22, pp.~12599--12610, 2022.

\bibitem{abhishek2020hybrid}
B.~Abhishek, S.~Ranjit, T.~Shankar, G.~Eappen, P.~Sivasankar, and A.~Rajesh, ``Hybrid pso-hsa and pso-ga algorithm for 3d path planning in autonomous uavs,'' {\em SN Applied Sciences}, vol.~2, pp.~1--16, 2020.

\bibitem{8820322}
S.-Y. Shin, Y.-W. Kang, and Y.-G. Kim, ``Automatic drone navigation in realistic 3d landscapes using deep reinforcement learning,'' in {\em 2019 6th International Conference on Control, Decision and Information Technologies (CoDIT)}, pp.~1072--1077, 2019.

\bibitem{azar2021drone}
A.~T. Azar, A.~Koubaa, N.~Ali~Mohamed, H.~A. Ibrahim, Z.~F. Ibrahim, M.~Kazim, A.~Ammar, B.~Benjdira, A.~M. Khamis, I.~A. Hameed, {\em et~al.}, ``Drone deep reinforcement learning: A review,'' {\em Electronics}, vol.~10, no.~9, p.~999, 2021.

\bibitem{hodge2021deep}
V.~J. Hodge, R.~Hawkins, and R.~Alexander, ``Deep reinforcement learning for drone navigation using sensor data,'' {\em Neural Computing and Applications}, vol.~33, no.~6, pp.~2015--2033, 2021.

\bibitem{9767604}
H.~Fahimi, S.~H. Mirtajadini, and M.~Shahbazi, ``A vision-based guidance algorithm for entering buildings through windows for delivery drones,'' {\em IEEE Aerospace and Electronic Systems Magazine}, vol.~37, no.~7, pp.~32--43, 2022.

\bibitem{do2016differential}
M.~P. Do~Carmo, {\em Differential geometry of curves and surfaces: revised and updated second edition}.
\newblock Courier Dover Publications, 2016.

\bibitem{bishop1975there}
R.~L. Bishop, ``There is more than one way to frame a curve,'' {\em The American Mathematical Monthly}, vol.~82, no.~3, pp.~246--251, 1975.

\bibitem{choi2002euler}
H.~I. Choi and C.~Y. Han, ``Euler--rodrigues frames on spatial pythagorean-hodograph curves,'' {\em Computer Aided Geometric Design}, vol.~19, no.~8, pp.~603--620, 2002.

\bibitem{shani1984splines}
U.~Shani and D.~H. Ballard, ``Splines as embeddings for generalized cylinders,'' {\em Computer Vision, Graphics, and Image Processing}, vol.~27, no.~2, pp.~129--156, 1984.

\bibitem{chirikjian2013framed}
G.~S. Chirikjian, ``Framed curves and knotted dna,'' {\em Biochemical Society Transactions}, vol.~41, no.~2, pp.~635--638, 2013.

\bibitem{karagiannis2011vehicular}
G.~Karagiannis, O.~Altintas, E.~Ekici, G.~Heijenk, B.~Jarupan, K.~Lin, and T.~Weil, ``Vehicular networking: A survey and tutorial on requirements, architectures, challenges, standards and solutions,'' {\em IEEE communications surveys \& tutorials}, vol.~13, no.~4, pp.~584--616, 2011.

\bibitem{7093187}
J.~Zheng and Q.~Wu, ``Performance modeling and analysis of the ieee 802.11p edca mechanism for vanet,'' {\em IEEE Transactions on Vehicular Technology}, vol.~65, no.~4, pp.~2673--2687, 2016.

\bibitem{hasrouny2017vanet}
H.~Hasrouny, A.~E. Samhat, C.~Bassil, and A.~Laouiti, ``Vanet security challenges and solutions: A survey,'' {\em Vehicular Communications}, vol.~7, pp.~7--20, 2017.

\bibitem{theodore2002wireless}
S.~R. Theodore {\em et~al.}, ``{Wireless Communications: Principles and Practice},'' in {\em Upper Saddle River}, Prentice Hall, 2002.

\bibitem{Moravejosharieh:Willig:15}
A.~Moravejosharieh and A.~Willig, ``{Mutual Interference in Large Populations of Co-Located IEEE 802.15.4 Body Sensor Networks -- A Sensitivity Analysis},'' {\em {Elsevier Computer Communications}}, vol.~81, pp.~86 -- 96, May 2016.

\bibitem{Willig:25}
A.~Willig, ``{A Sensitivity Analysis of the DCP/Vardis Protocol for Coordinating Drone Swarms},'' in {\em Proc. IEEE 2025 Vehicular Networking Conference (VNC)}, (Porto, Portugal), June 2025.

\bibitem{Myers:Montgomery:02}
R.~Myers and D.~Montgomery, {\em Response Surface Methodology}.
\newblock John Wiley and Sons,Inc, 2002.

\bibitem{Jain:91}
R.~Jain, {\em The Art of Computer Systems Performance Analysis -- Techniques for Experimental Design, Measurement, Simulation, and Modeling}.
\newblock Wiley Professional Computing, New York, Chichester: John Wiley and Sons, 1991.

\end{thebibliography}

\end{document}